\documentclass[fleqn,11pt]{cas-dc} 

\usepackage{amsfonts,amssymb}
\usepackage{amsmath}
\usepackage{rotating}
\usepackage{pdflscape,array,booktabs}
\usepackage{tablefootnote}
\usepackage{threeparttable}
\newcolumntype{P}[1]{>{\raggedright}p{#1}}
\newlength{\ninetabcol}
\newlength{\eighttabcol}
\newlength{\sixtabcol}
\newlength{\fivetabcol}
\newcommand{\tn}{\tabularnewline}

\usepackage[authoryear]{natbib}

\def\tsc#1{\csdef{#1}{\textsc{\lowercase{#1}}\xspace}}
\tsc{WGM}
\tsc{QE}

\begin{document}
\let\WriteBookmarks\relax
\def\floatpagepagefraction{1}
\def\textpagefraction{.001}

\shorttitle{Constraining the Origin of Mars via Simulations of Multi-Stage Core Formation}    

\shortauthors{Nathan et al.}  

\title [mode = title]{Constraining the Origin of Mars with Simulations of Multi-Stage Core Formation}  



%

\author[1]{Gabriel Nathan}[type=author,orcid=0000-0002-4261-6677]

\cormark[1]


\ead{nathanga@msu.edu}



\address[1]{Dept. of Earth \& Environmental Sciences, Michigan State University, East Lansing, MI 48824, USA}

\author[2]{David C. Rubie}[type=author,orcid=0000-0002-3009-1017]





\address[2]{Bayerisches Geoinstitut, University of Bayreuth, D-95490 Bayreuth, Germany}

\author[1]{Seth A. Jacobson}[type=author,orcid=0000-0002-4952-9007]            




\begin{abstract}
It remains an elusive goal to simultaneously model the astrophysics of Solar System accretion while reproducing the mantle chemistry of more than one inner terrestrial planet.
Here, we used a multistage core-mantle differentiation model based on \citet{rubie2011heterogeneous,rubie2015accretion} to track the formation and composition of Earth and Mars in various Grand Tack formation simulations.
Prior studies showed that in order to recreate Earth’s mantle composition, it must grow first from reduced (Fe-metal rich and O-poor) building blocks and then from increasingly oxidized (FeO rich) material.
This accretion chemistry occurs when an oxidation gradient exists across the disk so that the innermost solids are reduced and increasingly oxidized material is found at greater heliocentric distances.
For a suite of Grand Tack simulations, we investigated whether Earth and Mars can be simultaneously produced by the same oxidation gradient.
Our model did not find an oxidation gradient that simultaneously reproduces the mantle compositions of Earth and Mars.
Due to its small mass and rapid formation, the formation history of Mars-like planets is very stochastic which decreases the likelihood of compatibility with an Earth-producing oxidation gradient in any given realization. 
To reconcile the accretion history and ideal chemistry of the Mars-like planet with the oxidation gradient of an Earth-producing disk, we determined where in the Earth-producing disk Mars must have formed.
We find that the FeO-rich composition of the Martian mantle requires that Mars' building blocks must originate exterior to 1.0 astronomical units (AU). 
\end{abstract}




\maketitle

\section{Introduction}
In the past two decades, attempts to numerically recreate the astrophysical properties of the inner Solar System have made tremendous progress. 
N-body simulations of Solar System formation are now able to reproduce the structure of the inner planets and match many dynamical constraints \citep{chambers2001making, kokubo1998oligarchic}. 
Early models of Solar System formation were able to reproduce the four inner planets’ semi-major axes and two of the four planets' approximate masses (Earth and Venus). 
However, these early simulations often encountered a persistent difficulty in simulating an accurately sized Mars, known as the ``small Mars problem''; they routinely created a Mars-like planet with a mass that was too large \citep[e.g.][]{raymond2009building}.
The sharply decreasing mass ratio from Earth to Mars to the asteroid belt was one of the hardest astrophysical constraints of the inner Solar System to match, but there are now several proposed mechanisms that can recreate the mass-orbit distribution of the inner Solar System.
Among these are a truncated planet-forming disk \citep{hansen2009formation,lykawka2019constraining}, the Grand Tack model \citep{walsh2011low}, a steep surface density model \citep{izidoro2014terrestrial}, slow runaway growth \citep{walsh2019planetesimals}, early planet instability \citep{clement2018mars}, and an ad hoc pebble accretion scenario \citep{johansen2021pebble}. 
The many possible solutions to the small Mars problem represent significant progress in the field from when the problem was definitively described in \citet{raymond2009building}.
In order to distinguish between the many possible Solar System formation scenarios that are compatible with astrophysical constraints, we must next investigate the compatibility of these theories with the known geochemistry and physical properties of the inner terrestrial planets.

In this study, we focus our investigation on the Grand Tack model, which is among the first terrestrial planet formation scenarios to solve the small Mars problem \citep{walsh2011low,jacobson2014lunar} and has been shown to reproduce many aspects of Earth's chemistry and possibly that of Venus' \citep{jacobson2014lunar,rubie2015accretion,rubie2016highly}.
The Grand Tack scenario hypothesizes that planet-gas interactions in the early protoplanetary disk resulted in a 2:3 mean-motion resonance between Saturn and Jupiter that causes Jupiter to migrate outward \citep{masset2001reversing, morbidelli2007dynamics} after a period of inward migration.
Jupiter's migration shapes the inner disk into a narrow annulus within which the inner planets grow amidst a great deal of radial mixing \citep{hansen2009formation}.
Notably, in this scenario, the area of the disk surrounding Mars is cleared out and Mars grows in an empty zone that prematurely terminates its growth, successfully addressing the small Mars problem \citep{walsh2011low}.
Astrophysical N-body simulations of the Grand Tack scenario show rapid and stunted growth histories of Mars-like planets \citep{walsh2011low,jacobson2014lunar} that are largely compatible with Hf-W evidence for Mars' rapid formation \citep{nimmo2007mars,dauphas2011hf, brennan2021timing} and the extended formation history of Earth \citep{walker2009highly,jacobson2014highly}.

However, the Grand Tack is not the only viable model of planet formation nor even necessarily the most correct model. 
The movement of Jupiter in the Grand Tack scenario results in significant radial mixing of the inner disk, allowing bodies to be sourced from material that originated a great distance away from their ultimate orbit. 
Such extreme radial mixing in this scenario is potentially problematic to the Grand Tack's ability to reproduce the existence of isotopic gradients in the disk as a function of heliocentric distance \citep{brasser2017cool, woo2018curious} such as is observed with Cr  \citep{yamakawa2010chromium}. 

Despite this, our results are likely applicable to other planetary formation models, in addition to just the Grand Tack. 
Rather than focusing on validating the specific mechanics of the Grand Tack scenario, our study relies on broad trends in the oxidation states of Earth and Mars' mantles to ultimately suggest that Mars is built more locally while Earth samples the entire inner disk.
Such a conclusion is compatible with other formation scenarios such as the early instability scenario \citep{clement2018mars} or those that suggest the isotopic compositions of Earth and Mars may require less radial mixing than would be expected in the Grand Tack formation scenario \citep{izidoro2014terrestrial, raymond2017empty, mah2021isotopically}. 
Another alternative is a pure pebble accretion scenario \citep{johansen2021pebble} where Earth and Mars would have grown from the same pebble flux.
In such a scenario, it has been proposed that the relative mass ratio of an initial embryo with Ureillitic composition to the mass accreted as pebbles with CI carbonaceous chondrite composition determines the bulk composition of the accreted planet \citep{schiller2018isotopic}.

Multiple studies have examined the relationship between Mars' chemical and isotopic composition, potential dynamical formation scenarios, and a genetic link (or lack thereof) to Earth's building blocks \citep[e.g.][]{bond2010making,brasser2013formation, brasser2017colossal, brasser2017cool, brasser2018jupiter,dauphas2011hf,burbine2004determining,fitoussi2016building, brennan2021timing}.
By combining dynamical modeling with chemistry, \citet{bond2010making} first showed that dynamical models (in particular, those of \citet{o2006terrestrial}) are compatible with producing the elemental compositions of the inner planets. 
Isotopic evidence from the Hf/W system supports a rapid accretion and core formation timescale for Mars, which is consistent with dynamical formation models as well as with Mars' status as a 'stranded planetary embryo' \citep{dauphas2011hf,jacobson2014lunar,brennan2021timing}.
Isotopic mixing models reach differing conclusions as to whether Earth and Mars share similar building blocks, with some showing a link between the two bodies' progenitor material \citep{fitoussi2016building} and others supporting distinct reservoirs forming the two bodies \citep{burbine2004determining}.
\citet{fitoussi2016building} used a Monte-Carlo inversion model to deduce that Earth and Mars must share nearly the same building block material, though this study did not model dynamical interactions between growing planetary bodies.
Dynamical models of Mars' formation show too much variability and do not tightly constrain Mars' accretion history, but its distinct isotopic composition points towards an origin in the outer part of the inner Solar System \citep{brasser2017cool, brasser2018jupiter}. 
All of these studies advance the goal of understanding the connection between the chemical and isotopic composition of initial planetary building blocks, their accretion histories, and the final compositions of Earth and Mars.
However, the model described in \citet{rubie2015accretion} is unique in directly combining astrophysical N-body dynamics from \citep{jacobson2014lunar} with mantle and core elemental abundance evolution due to metal-silicate equilibration during planetary differentiation \citep{rubie2011heterogeneous}.

\citet{rubie2015accretion} used a joint astronomical-geochemical model to demonstrate the compatibility of multi-stage core formation with the Grand Tack model and to understand this formation scenario's effect on elemental abundances in the growing Earth.
Furthermore, they demonstrated that the Grand Tack scenario is capable of reproducing the bulk composition of a simulated Earth-like body when 1) the planetesimals and embryos that originate close to the Sun are highly reduced and those at greater heliocentric distances are increasingly oxidized and 2) the growing Earth-like planet experiences multiple core formation events.
Using a similar methodology, \citet{rubie2016highly} further showed that highly siderophile element (HSE) mantle abundances will increase throughout a multi-stage core formation because the HSE metal-silicate partition coefficients decrease at higher pressure and temperature conditions associated with core-formation.
Related models were used to investigate the effect of inefficient accretion on planetary compositional diversity \citep{cambioni2021effect} and the role of carbon in determining the bulk silicate composition of the Earth \citep{jennings2021metal}.

Here, we extend the methodology of \citet{rubie2015accretion} to investigate the origin of Mars within the protoplanetary disk.
We use N-body astrophysical simulations of the Grand Tack coupled with models of core-mantle differentiation to study the joint physical-chemical evolution of Earth, Mars, and the Solar System.
Specifically, for this work, we use this model to seek out the necessary conditions to reproduce Mars-like planets with appropriate mantle chemistry and core mass fractions.
\citet{rubie2015accretion} reproduced the bulk composition of Earth in the context of the Grand Tack scenario by fitting parameters associated with an Fe and Si oxidation gradient and an equilibration pressure factor.
\citet{rubie2015accretion} also examined the composition of Mars-like planets in those same simulations--but not directly fitting for them, and they concluded that the match to Mars was poor because it did not originate in an ideal location in the disk.
They suggested Mars must come from a specific location in the protoplanetary disk to reproduce its mantle FeO concentration, and in this study we used a more rigorous process to determine where in the disk Mars must originate. 
Here, we used a very similar model to find a set of fitted parameters (Fe and Si oxidation gradient and pressure factor) that could simultaneously reproduce Earth and Mars.

Simulating the formation of Mars presents significant challenges; it has a much smaller mass than Earth and a much more rapid formation.  
Because orbital dynamics and accretion in a disk are stochastic processes, there are multiple ways in which any individual planetary body may acquire its mass and orbital parameters.
Its constituent chemical components may be sourced from any region in the protoplanetary disk and still reproduce Mars' astrophysical constraints: mass and orbit. 
Mars' small size and rapid formation means it must be built from a few locally sourced embryos with only a limited planetesimal mass contribution.
This limits the  stochastic nature of its origin relative to Earth. 
In this paper, we investigate whether this stochastic growth history can be made consistent with Mars' bulk chemistry in the framework of the Grand Tack scenario.

%
\begin{table*}[width=\textwidth,cols=8,pos=t]
\caption{Astrophysical N-body initial conditions of the 6 Grand Tack simulations used in this study, from \citep{jacobson2014lunar}.} \label{tbl1}
\begin{tabular*}{\textwidth}{@{}P{\eighttabcol}P{\eighttabcol}P{\eighttabcol}P{\eighttabcol}P{\eighttabcol}P{\eighttabcol}P{\eighttabcol}P{\eighttabcol}@{}}
\toprule
Simulation & Embryo source region (AU) & Planetesimal source regions (AU) & Mass of embryos (M$_e$)  & Mass of planetesimals (M$_p$)  & Number of embryos & Number of planetesimals & Mass adjustment factor\tn \midrule
4:1-0.25-7 & 0.7-3.0 & 0.7-3.0 & 0.025 & 3.9  $\times 10^{-4}$ & 170 & 2783 & 1.04\tn
 &  & 6.0-9.5 & & 4.8 $\times 10^{-5}$ & &  1563 &  \tn 
4:1-0.5-8  & 0.7-3.0 & 0.7-3.0 & 0.05 & 3.9 $\times 10^{-4}$ & 87 & 2836 & 1.06\tn
 &  & 6.0-9.5 & & 4.8 $\times 10^{-5}$ & &  1563 &  \tn 
8:1-0.25-2 & 0.7-3.0 & 0.7-3.0 & 0.025 & 3.9 $\times 10^{-4}$ & 213 & 1750 & 0.91\tn
 &  & 6.0-9.4 & & 1.2 $\times 10^{-4}$ & &  500  & \tn %
8:1-0.8-18 & 0.7-3.0 & 0.7-3.0 & 0.08 & 3.9 $\times 10^{-4}$ & 68 & 1750 & 0.94\tn
 &  & 6.0-9.4 & & 4.8 $\times 10^{-5}$ & &  500  & \tn %
i-4:1-0.8-4 & 0.5-3.0 & 0.5-3.0 & 0.02-0.08 & 3.9 $\times 10^{-4}$ & 82 & 2260 & 1.01\tn
 &  & 6.0-9.4 & & 1.2 $\times 10^{-4}$ & &  500  & \tn %
i-4:1-0.8-6 & 0.5-3.0 & 0.5-3.0 & 0.02-0.08 & 3.9 $\times 10^{-4}$ & 82 & 2260 & 0.92\tn
 &  & 6.0-9.4 & & 1.2 $\times 10^{-4}$ & &  500  & \tn %
\bottomrule
\end{tabular*}
\end{table*}
\section{Methods}
\label{section:methods}

\begin{table*}[width=\textwidth,cols=8,pos=t]
\caption{Final masses in terms of Earth mass, semi-major axes, and mass adjustment factors of Earth and Mars embryos in the 6 Grand Tack simulations used. Mass adjustment factor is applied to set the mass of the final Earth analog equal to one Earth mass, and is also applied to the mass of the Mars embryo.} \label{tbl_final_earth_mars}
\begin{tabular*}{\textwidth}{@{}P{\sixtabcol}P{\sixtabcol}P{\sixtabcol}P{\sixtabcol}P{\sixtabcol}P{\sixtabcol}@{}}
\toprule
Simulation & M$_{\text{Earth}}$ & $a_{\text{Earth}}$ & M$_{\text{Mars}}$ & $a_{\text{Mars}}$ & Mass adjustment factor\tn \midrule
4:1-0.25-7 & 0.9570 & 0.9072 & 0.2202 & 1.3631 & 1.04\tn
  
4:1-0.5-8  & 0.9368 & 0.9689 & 0.1025 & 1.6784 & 1.06\tn
 
8:1-0.25-2 & 1.1070 & 0.9356 & 0.1376 & 1.4753 & 0.91\tn
 
8:1-0.8-18 & 1.0734 & 1.1794 & 0.0868 & 1.7621 & 0.94\tn
 
i-4:1-0.8-4 & 0.9980 & 0.7899 & 0.2779 & 1.4612 & 1.01\tn
 
i-4:1-0.8-6 & 1.1014 & 0.8810 & 0.0523 & 1.8989 & 0.92\tn
 
\bottomrule
\end{tabular*}
\end{table*}

We simulate the growth and chemical evolution of the terrestrial planets following the methods used in \citet{rubie2015accretion}.
This model couples N-body astrophysical simulations of Solar System formation with a model of core-mantle differentiation that iteratively refines initial input parameters to produce a final bulk chemical composition of a planetary body matching the mantle composition of either Earth or Mars. 
The N-body simulations used in this study are of the Grand Tack Solar System formation scenario.
The Grand Tack refers to an inward migration, a ``tack'' due to the presence of Saturn, and a subsequent outward migration of Jupiter’s early orbit which results in a disruption and subsequent determination of the inner planets’ orbits and is a leading model of Solar System formation \citep{walsh2011low,jacobson2014lunar}.  
Six simulations are taken from a larger suite of such simulations presented in \citet{jacobson2014lunar} and their corresponding properties are described in Table \ref{tbl1}. 

These particular simulations were chosen because, out of the set of simulations presented in \citet{jacobson2014lunar}, they produced particularly good approximations of the masses and semi-major axes of Earth and Mars, as shown in Table \ref{tbl_final_earth_mars}.
All of the Mars analogs except in Simulation i-4:1-0.8-6 have a mass greater than the lower cutoff defined in \citet{jacobson2014lunar} to classify a Mars analog. 
(To be classified as a Mars analog in \citet{jacobson2014lunar}, the body must exceed a mass of 0.0535 M$_{\bigoplus}$). 
We include the Mars analog in Simulation i-4:1-0.8-6 because it is very close to the cutoff (0.0523 M$_{\bigoplus}$ compared to 0.0535 M$_{\bigoplus}$, see Table \ref{tbl_final_earth_mars}) and because it is one of the simulations used in \citet{rubie2015accretion}. 
All simulations are the same as those used in \citet{rubie2015accretion}, with the exception of Simulation 8:1-0.8-18. 
Simulation 8:1-0.8-18 (Run 18) is from a suite of simulations with the same initial conditions that were used in \citet{rubie2015accretion} (in that case, Simulation 8:1-0.8-8 (Run 8) was used).
We chose Run 18 instead of Run 8 for this study because its mass ratio between the Earth and Mars analog is more accurate than that in Run 8.

Key information is conveyed by shorthand in the name of each simulation: "i" indicates that embryo mass increases with increasing heliocentric distance in the embryo source region, the ratio (4:1 or 8:1) indicates the ratio of total embryo mass to total planetesimal mass, the decimal indicates the initial mass of the largest embryo, and the final number indicates the run number of the simulation within the suite of simulations that has these initial conditions. 
The simulations chosen for this paper have higher initial embryo masses because then Mars-like planets are assembled fast enough to be consistent with evidence from the Hf-W radiometric system \citep{Jacobson2014d,Nimmo2007b,Dauphas2011}.
In other words, Mars is a stranded embryo and very likely indicative of how large embryos grew during runaway planetesimal growth or due to pebble accretion.
The mass adjustment factor column of Table \ref{tbl1} reports the scaling factor, applied to all masses in the simulation, used to make the final Earth-like planet mass equal to one Earth mass.
A notable difference between the simulations: for simulations 8:1-0.25-2 and 8:1-0.8-18, the ratio between total embryo mass and total planetesimal mass is 8:1 and for all other simulations, this ratio is 4:1.
All of the individual embryo masses in a given simulation are the same (and much greater than that of the planetesimals) unless denoted by an "i", as in simulations i-4:1-0.8-4 and i-4:1-0.8-6 where the individual embryo masses in the simulation increase as a function of semi-major axis for the range given in the column labeled "Mass of embryos (M$_e$)".

Each simulation is initialized with a disk of embryos (which interact with all particles in the simulation) and planetesimals (which interact only with the embryos and the Sun but not each other) in the inner disk--a standard computational technique in astrophysical N-body simulations. 
Because of the stochastic nature of N-body simulations, the dynamical evolution of these embryos may vary while still reproducing the masses and semi-major axes of the inner planets.
With each collision between embryos, the simulated planets grow more massive and evolve chemically due to core formation. 

Following an impact between simulated planetary embryos, a core-mantle differentiation event occurs.
In the reported simulations, the entire metal core of the projectile body equilibrates with a fraction of the silicate mantle of the target body. 
This fraction is determined by hydrodynamical laboratory experiments \citep{deguen2011experiments}, which model the entrainment of silicate liquid in the plume surrounding the core of the impactor as it descends through the magma ocean.
Metal-silicate equilibration occurs within this zone of mixed silicate liquid and molten metallic droplets.
The denser core-forming liquids sink through the mantle and merge with the target's core.
The silicate melt which equilibrated with the core-forming liquids is then mixed via convection with the non-equilibrating fraction of the mantle, including the accreted projectile mantle.

To reproduce the mantle elemental abundances of Earth-like or Mars-like planets, a set of initially unknown free parameters are varied in the model until the desired mantle composition is produced. 
These parameters describe a oxidation gradient in the protoplanetary disk that sets the initial compositions of the planetesimals and embryos.
A parameter also determines the P-T conditions of the metal-silicate equilibration events.
All of these parameters are described in more detail below.
By adjusting these free parameters to minimize a chi-squared metric assessing the match between modeled and measured mantle abundances of siderophile and lithophile elements, the model demonstrates its ability to 1) reproduce desired planetary compositions and 2) do so with a plausible set of conditions represented by each fitted parameter.
The exact value of each parameter during the formation of Earth or Mars cannot be directly known, but we can assess the reasonableness of the fitted parameters after-the-fact.
The strength of this method is that by not forcing specific values for the parameters and instead letting them float as free parameters, we can infer the ideal initial conditions by continually re-adjusting until a best-fit minimum chi-squared scenario is achieved. 

\begin{figure}
    \centering
    \includegraphics[width=\columnwidth]{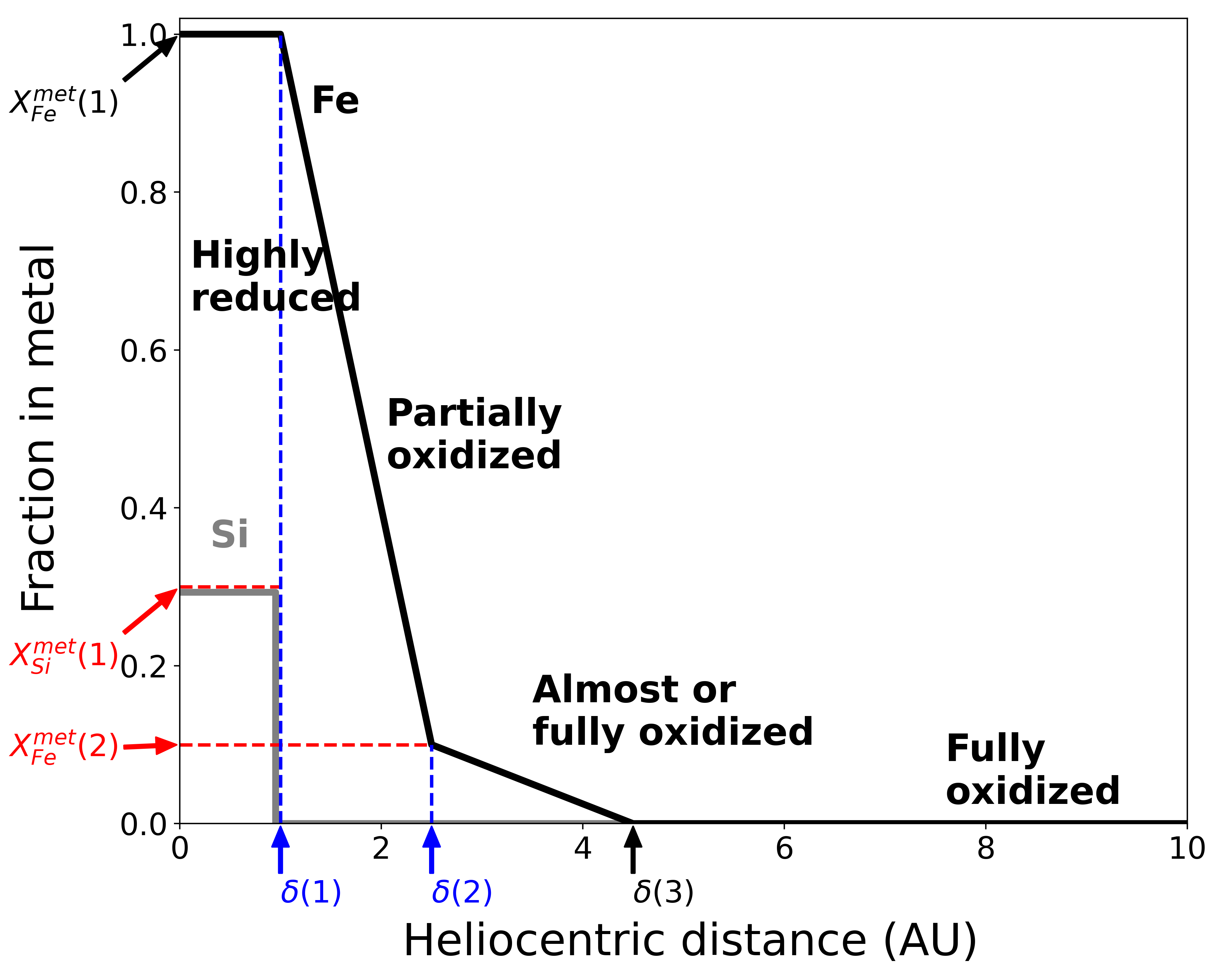} 
    \caption{Example Fe and Si oxidation gradients of starting primitive bodies, adapted from \citet{rubie2015accretion}.
    The black line shows the fraction of Fe in metal as a function of increasing heliocentric distance.
    The Fe oxidation state of embryos changes from highly reduced to fully oxidized as heliocentric distance increases, but the exact shape of the gradient is simulation-dependent.
    The Si oxidation state, in gray, is a step function that is simulation-dependent.
    The free parameters that determine fraction of Fe and Si in metal ($X^{met}_{Si}$(1) and $X^{met}_{Fe}$(2), in red) at different heliocentric distances ($\delta$(1) and $\delta$(2), in blue) are shown along the axes.
    $X^{met}_{Fe}$(1) $\delta$(3), in black, are fixed parameters.
    \label{fig:oxidation}}
\end{figure}

The parameters we explore control both the oxidation state of the embryos and planetesimals in the initial protoplanetary disk and the conditions of metal-silicate equilibration. 
The five parameters are as follows: $f_p$, $\delta$(1), $\delta$(2), $X^{met}_{Fe}$(2), and $X^{met}_{Si}$(1). 
$f_p$ is a constant proportionality factor that is multiplied by the core-mantle boundary (CMB) pressure of the impacted growing embryo to determine the pressure for each metal-silicate equilibration event.
The equilibration temperature is the mid-point between the solidus and liquidus of peridotite at the corresponding equilibration pressure. 
The other four parameters control the initial oxidation states of the embryos and planetesimals in the protoplanetary disk, as shown in Figure \ref{fig:oxidation}.
The free parameters $\delta$(1) and $\delta$(2) represent heliocentric distances in the disk at which the embryos' and planetesimals' respective fractions of the bulk Fe in metal are $X^{met}_{Fe}$(1) and $X^{met}_{Fe}$(2). 
$X^{met}_{Fe}$(1) is fixed to equal 1.0 and $X^{met}_{Fe}$(2) is a free parameter that can vary between 0 and 1.
$X^{met}_{Si}$(1) is a free parameter that represents the embryos' and planetesimals' initial fraction of bulk Si in metal between the heliocentric distances of 0 AU and $\delta$(1).
At distances greater than $\delta$(1), $X^{met}_{Si}$(1) is equal to zero.
Exterior to a third (fixed) heliocentric distance, $\delta$(3) = 4.5 AU, all embryos and planetesimals are fully oxidized and additionally are H$_2$O-bearing. 
This is similar to the free parameters used in \citet{jennings2021metal}, where this model is described in greater detail.
In the rest of this paper, we will refer to this set of four parameters, not including $f_p$, as an oxidation gradient.

We use a least squares minimization algorithm to adjust all five of the free parameters to find best-fit conditions to reproduce particular planetary mantle compositions.
This algorithm uses a nonlinear multidimensional minimization routine called the downhill simplex method (\citet{nelder1965simplex}) to explore a multidimentional parameter space in order to find a global minimum on a rough chi-square landscape with potentially many local minima. 
We run this algorithm for thousands of initial sets of initial conditions, called simplexes, to ensure we have found the best fit parameters corresponding closest to a true global minimum in the chi-square surface.
By using this algorithm on many sets of simplexes and comparing the best fit values, we are able to quantify an uncertainty in the parameters that define the true global minimum.

An uncertainty estimate of the modeled mantle composition is derived from standard error propagation of the 1-sigma measurement error from laboratory partition coefficients for each element through the numerical model, similar to \citet{rubie2015accretion}. 
For the best fit parameters, we calculate an upper and lower bound on partitioning behavior and corresponding mantle abundance.
In Tables \ref{tblbothfittaylor} through \ref{tblmarsfityoshizaki}, we report the average of the difference of upper and lower bound compositions from the best-fit abundance as an estimate of uncertainty of a each simulated mantle elemental abundance.

\begin{table*}[width=\textwidth,cols=8,pos=h]
    \caption{Fitting the model to Earth and resulting best fit parameters described in Section \ref{section:methods}. Resulting best fit parameters and chi squared estimated errors are shown for fitting to Earth's composition. The results below are directly comparable to \citet{rubie2015accretion} and do not reflect the use of a partition coefficient switch for Ni and Co at pressures of <5 GPa.
    \label{tblno5gpaswitchparam}}
\begin{tabular*}{\textwidth}{@{}P{\eighttabcol}P{\eighttabcol}P{\eighttabcol}P{\eighttabcol}P{\eighttabcol}P{\eighttabcol}P{\eighttabcol}P{\eighttabcol}@{}}
\toprule
Simulation & Fit to & $f_p$ & $\delta$(1) & $\delta$(2) & $X^{met}_{Si}$(1) & $X^{met}_{Fe}$(2) & $\chi^2_{red}$ \tn
\midrule
4:1-0.25-7  & Earth & 0.56 $\pm$ 0.03 & 0.99 $\pm$ 0.14 & 1.42 $\pm$ 0.81 & 0.24 $\pm$ 0.12 & 0.53 $\pm$ 0.41& 2.9 \tn
4:1-0.5-8  & Earth & 0.68 $\pm$ 0.03 & 1.18 $\pm$ 0.24 & 1.18 $\pm$ 0.92 & 0.11 $\pm$ 0.08 & 0.49 $\pm$ 0.4  & 2.1 \tn
8:1-0.25-2  & Earth & 0.66 $\pm$ 0.03 & 1.21 $\pm$ 0.14 & 1.25 $\pm$ 0.33 & 0.13 $\pm$ 0.03 & 0.52 $\pm$ 0.2  & 1.1 \tn
8:1-0.8-18 & Earth & 0.7 $\pm$ 0.06 & 1.31 $\pm$ 0.27 & 1.31 $\pm$ 0.45 & 0.11 $\pm$ 0.05 & 0.3 $\pm$ 0.28  & 3.3 \tn
i-4:1-0.8-4 & Earth & 0.7 $\pm$ 0.03 & 1.03 $\pm$ 0.09 & 1.05 $\pm$ 0.34 & 0.14 $\pm$ 0.04 & 0.52 $\pm$ 0.28 & 1.3 \tn
i-4:1-0.8-6 & Earth & 0.66 $\pm$ 0.03 & 1.12 $\pm$ 0.16 & 1.15 $\pm$ 0.35 & 0.13 $\pm$ 0.05 & 0.56 $\pm$ 0.27 & 1.5 \tn
\bottomrule
\end{tabular*}
\end{table*}
\subsection{Planetary Compositions}
\label{section:composition}

We direct the model to find best fit parameters to reproduce the mantles of Earth \citep{palme2014cosmochemical} and Mars \citep{taylor2013bulk,yoshizaki2020composition}, either fitting jointly or individually.
The model tracks the evolution of major lithophile (Mg, Ca, Al, Na) and moderately siderophile elements (Fe, Si, O, Ni, Co, V, Cr, Nb, and Ta).
We calculate a chi-square metric by comparing the following elements in the modeled mantles to measured values: Mg, Fe, Si, Ni, Co, V, Cr, Nb, and Ta.
This subset of elements includes a representative lithophile (Mg) and notable siderophiles (Fe, Si, Ni, Co, V, Cr, Nb, Ta). 
Oxygen is not used as a constraint because it is controlled by the free parameters described in Section \ref{section:methods}, and the other lithophiles in the model (Ca, Al, Na) are not used as constraints because their behavior is very similar to Mg.
We additionally constrain Earth's composition by its Nb/Ta ratio from \citet{munker2003evolution}.
\citet{rubie2011heterogeneous} included the Nb/Ta ratio as a separate constraint relative to the Nb and Ta absolute mantle abundances because that ratio is better known than the absolute abundances for the Earth.
While we continue to use this ratio constraint for Earth, we do not use it for Mars since it is not better known than the individual mantle abundances of Nb or Ta.
Furthermore, we only considered Ta as a constraint for Earth and not for Mars.
There is significant disagreement in the literature about how siderophile Ta is at the low pressures relevant to Mars' core formation \citep{mann2009evidence, corgne2008metal}.
This uncertainty complicates any attempt to fit to the abundance of Ta in simulated Mars-like planets.

For the composition of Mars' mantle, we fit the model to two separate estimates; one from \citet{taylor2013bulk} and one from \citet{yoshizaki2020composition}.
\citet{taylor2013bulk} follows the methods of \citet{wanke1994chemistry} and closely resembles their Martian mantle compositional estimate.
This estimation assumes the proportions of major elements in Mars reflects that of CI chondrites, and that CI chondrites are a fundamental building block of Mars.
\citet{taylor2013bulk} does not provide an abundance of Vanadium in Mars' mantle, however by following the same methods as \citet{taylor2013bulk}, we estimated a V abundance of 105 ppm with 5\% uncertainty assuming it behaves similarly as other refractory lithophile elements and a CI chondrite V abundance \citet{palme2014cosmochemical}.
Instead of using CI chondrite composition as a starting point, \citet{yoshizaki2020composition} combines Martian meteoritic composition and spacecraft observations to estimate the Martian bulk composition.
To first order, the bulk silicate compositions of Mars (BSM) in \citet{yoshizaki2020composition} and \citet{taylor2013bulk} are very similar; in each estimation of Mars' composition, Mars' mantle has a higher FeO content than that of Earth's mantle.
However, the \citet{taylor2013bulk} estimate of Mars' mantle FeO content is higher than in \citet{yoshizaki2020composition} (18.14 and 14.67 wt\%, respectively). 
As a result of its lower FeO content, the \citet{yoshizaki2020composition} composition is enriched in all non-Fe elements to which we fit our model.
We run our model to produce a Mars analog that resembles the mantle compositions of either \citet{yoshizaki2020composition} or \citet{taylor2013bulk}, starting from a disk of planetary embryos with non-volatile elements in chondritic relative proportions. 

The embryos and planetesimals in our model have compositions in CI chondritic relative abundances of non-volatile elements \citep{palme2014cosmochemical}, but vary as a function of heliocentric distance due to oxygen fugacity conditions described by free parameters. 
These parameters control the abundance of O, which controls the redox state of Si and Fe. 
We also enhance refractory element abundances in the initial bodies to account for the fact that the Al/Mg ratio of the Earth's mantle is significantly higher than the Al/Mg ratio of CI chondritic material (see \citet{rubie2011heterogeneous} for further explanation).
The following refractory elements are enhanced by 22\% with respect to CI abundances: Al, Ca, Nb, Ta, while V is increased by 11\%.

\subsection{Modifications to the model compared to \citet{rubie2015accretion}}
\label{section:modifications}

As previously discussed, this model is based on \citet{rubie2015accretion}, but it is updated in two notable ways. 1) It uses a different method to find a best-fit set of parameters, and 2) it handles Ni and Co partitioning differently. 
These updates improved our ability to find the best fit parameters by searching parameter space more thoroughly and improved out ability to model Ni and Co metal-silicate equilibration at lower P-T conditions. 
We still recover similar results to those reported in \citet{rubie2015accretion}, although with small differences.

The first change to the model allows it to explore parameter space more thoroughly at a cost of being more computationally expensive. 
\citet{rubie2015accretion} calculated a best fit set of parameters by manually providing sets (i.e. simplexes) of starting parameters to the least squares minimization algorithm, as described above. 
Here, instead of experimenting with different initial simplexes by hand, we assembled 1000 simplexes spanning the possible parameter space of each free parameter.  
Then, we minimized the reduced $\chi^2$ of every initial simplex using a least squares minimization algorithm.
The parameters defining the simplex with the lowest reduced $\chi^2$ was taken as our estimate of the global minimum and our best-fit solution.
We then estimated the uncertainty around this best-fit by finding the subset of simplexes that returned reduced $\chi^2$ values within one standard deviation of the best-fit (i.e. twice the reduced $\chi^2$ value).
From that subset of simplexes, we then averaged the minimum and maximum of each parameter value to obtain an estimate of the parameter uncertainty.
The benefit of this is twofold: we achieve a very thorough search of each parameter resulting in a better estimate of the global minimum, and we are also able to estimate the uncertainty of that global minimum.  
Due to this method of searching for a global minimum, we achieve a lower chi-square when fitting to the composition of the Earth than \citet{rubie2015accretion} did for each of the 6 simulations we use.
For this version of the model, which still handles Ni and Co the same way as \citet{rubie2015accretion}, Table \ref{tblno5gpaswitchparam} displays the best-fit and standard deviations of parameters, as well as the lowest achieved chi-squared value for each of the 6 Grand Tack scenarios.

\begin{table*}[width=\textwidth,cols=8,pos=h]
\caption{Best fit parameters of the 6 simulations for the free parameters $f_p$,  $\delta$(1), $\delta$(2), $X^{met}_{Si}$(1), and $X^{met}_{Fe}$(2) as described in Section \ref{section:methods}. Resulting best fit parameters and estimated errors are shown for fitting to Earth's primitive mantle composition. Results are shown for fitting to Earth and Mars' mantles jointly or solely, using either \protect \citet{yoshizaki2020composition} (Y) or \citet{taylor2013bulk} (T) values for Mars' mantle.}\label{tblparam}
\begin{tabular*}{\textwidth}{@{}P{\eighttabcol}P{\eighttabcol}P{\eighttabcol}P{\eighttabcol}P{\eighttabcol}P{\eighttabcol}P{\eighttabcol}P{\eighttabcol}@{}}
\toprule
Simulation & Fit to & $f_p$ & $\delta$(1) & $\delta$(2) & $X^{met}_{Si}$(1) & $X^{met}_{Fe}$(2) & $\chi^2_{red}$ \tn
\midrule
4:1-0.25-7  & Both (Y) & 0.57 $\pm$ 0.06 & 1.06 $\pm$ 0.2 & 1.08 $\pm$ 0.99 & 0.15 $\pm$ 0.18 & 0.65 $\pm$ 0.44 & 30.7 \tn
  & Both (T) & 0.57 $\pm$ 0.09 & 1.07 $\pm$ 0.22 & 1.08 $\pm$ 1.14 & 0.17 $\pm$ 0.33 & 0.63 $\pm$ 0.45  & 169.7 \tn
  & Earth    & 0.57 $\pm$ 0.03 & 0.99 $\pm$ 0.18 & 1.38 $\pm$ 0.8 & 0.24 $\pm$ 0.12 & 0.55 $\pm$ 0.41  & 2.9 \tn
  & Mars (Y) & 0.42 $\pm$ 0.16 & 0.72 $\pm$ 6.8 & 1.88 $\pm$ 2.48 & 0.35 $\pm$ 0.5 & 0.0003 $\pm$ 0.5  & 1.1 \tn
  & Mars (T) & 0.44 $\pm$ 0.09 & 0.82 $\pm$ 0.59 & 0.88 $\pm$ 1.14 & 0.11 $\pm$ 0.5 & 0.52 $\pm$ 0.32  & 1.5 \tn
\tn
4:1-0.5-8  & Both (Y) & 0.69 $\pm$ 0.07 & 0.95 $\pm$ 0.14 & 0.96 $\pm$ 0.29 & 0.22 $\pm$ 0.16 & 0.84 $\pm$ 0.12  & 20.2 \tn
  & Both (T) & 0.68 $\pm$ 0.08 & 0.95 $\pm$ 0.12 & 0.95 $\pm$ 0.26 & 0.22 $\pm$ 0.17 & 0.85 $\pm$ 0.1  & 25.6 \tn
  & Earth    & 0.69 $\pm$ 0.03 & 1.17 $\pm$ 0.24 & 1.18 $\pm$ 0.94 & 0.11 $\pm$ 0.08 & 0.6 $\pm$ 0.42  & 2.2 \tn
  & Mars (Y) & 0.53 $\pm$ 0.15 & -6.09 $\pm$ 3.9 & 2.42 $\pm$ 1.8 & 0.22 $\pm$ 0.5 & 0.67 $\pm$ 0.5  & 1.2 \tn
  & Mars (T) & 0.22 $\pm$ 0.1 & 1.5 $\pm$ 5.84 & 2.3 $\pm$ 2.11 & 1. $\pm$ 0.5 & 0.34 $\pm$ 0.5  & 2.4 \tn
\tn
8:1-0.25-2  & Both (Y) & 0.67 $\pm$ 0.1 & 1.37 $\pm$ 0.36 & 1.37 $\pm$ 1.26 & 0.11 $\pm$ 0.17 & 0.39 $\pm$ 0.46  & 93.9 \tn
  & Both (T) & 0.61 $\pm$ 0.12 & 0.9 $\pm$ 0.45 & 0.9 $\pm$ 1.45 & 0.28 $\pm$ 0.24 & 0.85 $\pm$ 0.47  & 287.5 \tn
  & Earth    & 0.66 $\pm$ 0.03 & 1.28 $\pm$ 0.14 & 1.29 $\pm$ 0.33 & 0.12 $\pm$ 0.03 & 0.5 $\pm$ 0.2  & 1.0 \tn
  & Mars (Y) & 0.0007 $\pm$ 0.48 & 0.85 $\pm$ 4.88 & 1.2 $\pm$ 2.54 & 0.11 $\pm$ 0.5 & 0.00007 $\pm$ 0.5  & 0.8 \tn
  & Mars (T) & 0.47 $\pm$ 0.18 & 0.88 $\pm$ 2.42 & 0.99 $\pm$ 2.89 & 0.6 $\pm$ 0.5 & 0.28 $\pm$ 0.49  & 2.2 \tn
\tn
8:1-0.8-18 & Both (Y) & 0.68 $\pm$ 0.08 & 1.38 $\pm$ 0.18 & 1.77 $\pm$ 0.34 & 0.1 $\pm$ 0.06 & 0.19 $\pm$ 0.18  & 11.0 \tn
  & Both (T) & 0.66 $\pm$ 0.12 & 1.31 $\pm$ 0.22 & 1.77 $\pm$ 0.57 & 0.12 $\pm$ 0.13 & 0.2 $\pm$ 0.35  & 31.1 \tn
  & Earth    & 0.7 $\pm$ 0.06 & 1.31 $\pm$ 0.28 & 1.31 $\pm$ 0.45 & 0.11 $\pm$ 0.05 & 0.3 $\pm$ 0.29  & 3.5 \tn
  & Mars (Y) & 0.38 $\pm$ 0.16 & 1.43 $\pm$ 4.34 & 1.76 $\pm$ 2.17 & 0.00001 $\pm$ 0.5 & 0.02 $\pm$ 0.5  & 1.1 \tn
  & Mars (T) & 0.22 $\pm$ 0.06 & 1.52 $\pm$ 0.34 & 1.54 $\pm$ 0.75 & 0.37 $\pm$ 0.46 & 0.03 $\pm$ 0.35  & 1.6 \tn
\tn
i-4:1-0.8-4 & Both (Y) & 0.71 $\pm$ 0.06 & 0.91 $\pm$ 0.1 & 0.91 $\pm$ 0.18 & 0.18 $\pm$ 0.11 & 0.75 $\pm$ 0.13  & 31.0 \tn
  & Both (T) & 0.7 $\pm$ 0.1 & 0.89 $\pm$ 0.12 & 0.89 $\pm$ 0.23 & 0.17 $\pm$ 0.14 & 0.76 $\pm$ 0.14  & 67.8 \tn
  & Earth    & 0.7 $\pm$ 0.04 & 1.03 $\pm$ 0.09 & 1.03 $\pm$ 0.32 & 0.14 $\pm$ 0.04 & 0.53 $\pm$ 0.28  & 1.6 \tn
  & Mars (Y) & 0.03 $\pm$ 0.2 & -9.61 $\pm$ 14.67 & 1.46 $\pm$ 2.06 & 0.93 $\pm$ 0.5 & 0.79 $\pm$ 0.5  & 1.4 \tn
  & Mars (T) & 0.00005 $\pm$ 0.06 & 0.82 $\pm$ 30.42 & 1.22 $\pm$ 2.38 & 0.05 $\pm$ 0.5 & 0.64 $\pm$ 0.5  & 2.0 \tn
\tn
i-4:1-0.8-6 & Both (Y) & 0.67 $\pm$ 0.04 & 1.07 $\pm$ 0.06 & 1.08 $\pm$ 0.09 & 0.15 $\pm$ 0.05 & 0.7 $\pm$ 0.1 & 15.8 \tn
  & Both (T) & 0.66 $\pm$ 0.07 & 1.07 $\pm$ 0.09 & 1.07 $\pm$ 0.7 & 0.16 $\pm$ 0.08 & 0.68 $\pm$ 0.37 & 18.6 \tn
  & Earth    & 0.65 $\pm$ 0.04 & 1.15 $\pm$ 0.16 & 1.15 $\pm$ 0.34 & 0.12 $\pm$ 0.05 & 0.48 $\pm$ 0.26 & 1.8 \tn
  & Mars (Y) & 0.57 $\pm$ 0.16 & 0.47 $\pm$ 2.81 & 2.24 $\pm$ 2.14 & 0.11 $\pm$ 0.49 & 0.49 $\pm$ 0.5 & 1.2 \tn
  & Mars (T) & 0.35 $\pm$ 0.08 & 1.47 $\pm$ 5.44 & 1.67 $\pm$ 2.48 & 0.34 $\pm$ 0.5 & 0.16 $\pm$ 0.5 & 2.2 \tn
\bottomrule
\end{tabular*}
\end{table*}

Secondly, we have updated the way the model handles Ni and Co partitioning relative to \citet{rubie2015accretion}, by allowing the code to use a second set of low-pressure partition coefficients for Ni and Co.
This change has very little effect on the accretion of the Earth, as shown by the similar values in Table \ref{tblno5gpaswitchparam} and \ref{tblparam}, thereby preserving the results of \citet{rubie2015accretion}, where the partition coefficients for Ni and Co were estimated using fits to data from high pressure laboratory experiments that exceeded 5 GPa. 
For the formation of the Earth, this pressure regime is acceptable since equilibration pressures are almost always greater than 5 GPa.
However, in this work we model Mars, whose small mass means a significantly higher proportion of its equilibration events occur at less than 5 GPa.
To approximate low pressure equilibration more accurately, we have implemented a function that switches between functional fits to data from high and low pressure Ni and Co experiments, using data from \citet{kegler2008new} as refit by \citet{rubie2011heterogeneous} (see Supplementary Data of \citet{rubie2011heterogeneous}). 
Without this change, it is not possible to reproduce Mars' mantle abundances of Ni and Co.
With the exception of the results in Table \ref{tblno5gpaswitchparam}, all results displayed are from the version of the model that handles Ni and Co in this way.

\section{Results}
\label{section:results}

We used our model to investigate the oxidation gradients that best produce Earth and Mars in the following ways.
We fit the model to simultaneously reproduce the mantle compositions of Earth and Mars, the results of which are shown in Tables \ref{tblbothfittaylor} and \ref{tblbothfityoshizaki}.
We also fit the model to reproduce Earth and Mars individually, the results of which are shown in Tables \ref{tblearthfitwith5gpaswitch}, \ref{tblmarsfittaylor}, and \ref{tblmarsfityoshizaki}.
With both sets of these results, we reconciled the best-fit conditions that reproduce Mars with those that best reproduce the Earth, the results of which are shown in Figures \ref{fig:idealtaylor} and \ref{fig:idealyoshi}. 
We report the best-fit parameters and standard deviation for each of the 6 Grand Tack scenario simulations in Table \ref{tblparam} for models reproducing Earth and Mars, with both joint and individual fits.

\begin{table*}[width=\textwidth,cols=9,pos=t]
\caption{Results of attempting to reproduce Earth and Mars mantle compositions simultaneously, using \citet{taylor2013bulk} as an estimate for Mars' composition. Mantle abundances are shown in wt\% (or ppm/ppb where indicated) for Earth and Mars-like planets for all simulations. $\chi^2_{red}$ values ares high, indicating that this approach results in poor approximations of the compositions of both the Earth and Mars-like planets.} \label{tblbothfittaylor}
\begin{threeparttable}
\begin{tabular*}{\textwidth}{@{}P{\eighttabcol}P{\eighttabcol}P{\eighttabcol}P{\eighttabcol}P{\eighttabcol}P{\eighttabcol}P{\eighttabcol}P{\eighttabcol}@{}}
\toprule
& Simulations: & 4:1-0.25-7  & 4:1-0.5-8  & 8:1-0.25-2  & 8:1-0.8-18 & i-4:1-0.8-4 & i-4:1-0.8-6 \tn 
\midrule 
\underline{Earth} & Measured\tnote{a}  \tn 
FeO  & 8.1 (1.29) &   8.25 (0.02) & 8.08 (0.03) & 8.19 (0.05) & 8.09 (0.07) & 8.09 (0.03) & 8.1 (0.03) \tn 
SiO$_2$  & 45.4 (2.14)&   45.44 (0.1) & 45.32 (0.09) & 44.99 (0.07) & 45.47 (0.08) & 45.29 (0.09) & 45.2 (0.09)  \tn 
Ni (ppm) &  1860 (5)&  1823.2 (990.6) & 1715.2 (953.8) & 1368.0 (764.4) & 1526.6 (946.9) & 1692.4 (958.8) & 1735.8 (952.2)  \tn 
Co (ppm)  & 102 (5)&  110.2 (55.8) & 103.2 (60.0) & 83.6 (46.1) & 113.1 (57.7) & 102.8 (58.7) & 106.0 (60.7)  \tn 
Nb (ppb) &  595 (20)&   523.7 (24.0) & 554.9 (18.9) & 633.5 (15.8) & 578.0 (30.9) & 569.7 (16.7) & 571.9 (16.6)  \tn 
Ta (ppb) & 43 (5) &  39.8 (0.7) & 39.7 (0.7) & 40.4 (0.7) & 40.8 (0.5) & 40.1 (0.6) & 40.4 (0.5)  \tn 
Nb/Ta & 14.0 (0.3)\tnote{b} & 13.4 (0.8) & 14.2 (0.7) & 15.8 (0.7) & 14.4 (0.9) & 14.2 (0.6) & 14.3 (0.6) \tn
V (ppm)  & 86 (5)&  84.3 (18.6) & 93.4 (13.9) & 94.6 (13.8) & 84.6 (20.9) & 89.0 (16.1) & 88.5 (18.2)  \tn 
Cr (ppm)  & 2520 (10)&   2856.1 (298.1) & 3346.5 (269.7) & 3408.6 (255.3) & 2723.7 (349.3) & 3134.9 (275.6) & 2993.9 (304.8)  \tn

\tn
\underline{Mars} & Measured\tnote{c} \tn
FeO  & 18.14 (0.5) &  9.76 (0.02) & 18.15 (0.1) & 8.99 (0.02) & 16.91 (0.11) & 16.52 (0.1) & 17.95 (0.09)  \tn 
SiO$_2$  & 43.85 (1.0) & 46.39 (0.03) & 43.51 (0.02) & 48.73 (0.02) & 44.31 (0.02) & 43.8 (0.02) & 43.7 (0.02)  \tn 
Ni (ppm) &  330 (54.5) &  230.9 (146.2) & 609.4 (666.0) & 167.2 (75.1) & 621.3 (686.2) & 871.6 (672.8) & 541.8 (579.5)  \tn 
Co (ppm)  & 71 (12.5) & 34.7 (20.9) & 88.7 (81.0) & 24.0 (13.2) & 91.1 (80.6) & 99.4 (75.1) & 79.8 (74.7)  \tn 
Nb (ppb) &  501 (0.035) & 405.4 (12.2) & 655.7 (0.1) & 707.7 (1.6) & 666.7 (0.2) & 589.5 (2.3) & 653.7 (0.8)  \tn 
V (ppm)  & 105 (5.25) & 87.1 (15.0) & 114.4 (0.9) & 123.7 (4.2) & 116.0 (1.4) & 107.0 (4.3) & 114.3 (1.3)  \tn 
Cr (ppm)  & 4990 (210) &  3159.1 (249.8) & 4674.7 (94.7) & 4677.6 (242.0) & 4677.7 (119.4) & 4271.9 (130.0) & 4662.9 (100.5)  \tn

\tn
$\chi^2_{red}$ & Fit to both &  169.71 &  25.56 & 287.48 & 31.08 & 67.84 & 18.57 \tn

\bottomrule
\end{tabular*}
\begin{tablenotes}
      \item[a]{\citet{palme2014cosmochemical}}
      \item[b]{\citet{munker2003evolution}}
      \item[c]{\citet{taylor2013bulk}}
    \end{tablenotes}
\end{threeparttable}
\end{table*}

%
\begin{table*}[width=\textwidth,cols=9,pos=t]
\caption{Results of attempting to reproduce Earth and Mars mantle compositions simultaneously, using \citet{yoshizaki2020composition} as an estimate for Mars' composition. Mantle abundances are shown in wt\% (or ppm/ppb where indicated) for Earth and Mars-like planets for all simulations. $\chi^2_{red}$ values ares high, indicating that this approach results in poor approximations of the compositions of both the Earth and Mars-like planets for Earth and Mars-like planets for all simulations.} \label{tblbothfityoshizaki}
\begin{threeparttable}
\begin{tabular*}{\textwidth}{@{}P{\eighttabcol}P{\eighttabcol}P{\eighttabcol}P{\eighttabcol}P{\eighttabcol}P{\eighttabcol}P{\eighttabcol}P{\eighttabcol}@{}}
\toprule
& Simulations: & 4:1-0.25-7  & 4:1-0.5-8  & 8:1-0.25-2  & 8:1-0.8-18 & i-4:1-0.8-4 & i-4:1-0.8-6 \tn 
\midrule
\underline{Earth} & Measured\tnote{a} \tn
FeO  & 8.1 (1.29) & 8.12 (0.02) & 8.08 (0.03) & 8.1 (0.03) & 8.07 (0.07) & 8.08 (0.03) & 8.08 (0.07)   \tn 
SiO$_2$  & 45.4 (2.14) &45.78 (0.1) & 45.27 (0.09) & 45.38 (0.09) & 45.51 (0.08) & 45.16 (0.1) & 45.16 (0.08)   \tn 
Ni (ppm) &  1860 (5) & 1828.5 (993.0) & 1776.2 (975.8) & 1792.9 (852.7) & 1641.1 (984.8) & 1745.4 (985.9) & 1854.4 (942.1)   \tn 
Co (ppm)  & 102 (5) &  109.8 (55.7) & 105.5 (60.8) & 108.5 (49.7) & 117.9 (58.5) & 105.0 (59.8) & 110.2 (59.3)   \tn 
Nb (ppb) &  595 (20) & 536.8 (23.8) & 560.7 (19.4) & 593.7 (21.2) & 577.3 (32.6) & 571.1 (17.0) & 577.3 (16.8)   \tn 
Ta (ppb) & 43 (5) & 40.0 (0.6) & 39.9 (0.7) & 41.2 (0.4) & 40.9 (0.5) & 40.2 (0.7) & 40.5 (0.5)   \tn 
Nb/Ta & 14.0 (0.3)\tnote{b} & 13.4 (0.8) & 14.4 (0.8) & 14.5 (0.7) & 14.4 (1.0) & 14.3 (0.7) & 14.4 (0.6) \tn
V (ppm)  & 86 (5) & 86.4 (18.5) & 93.6 (14.0) & 86.5 (21.6) & 84.5 (21.2) & 88.6 (16.2) & 88.8 (18.2)   \tn 
Cr (ppm)  & 2520 (10) & 2914.1 (304.2) & 3346.7 (271.4) & 2735.6 (356.6) & 2708.6 (353.1) & 3117.1 (276.1) & 3003.5 (305.5)   \tn 

\tn
\underline{Mars} & Measured\tnote{c} \tn
FeO  & 14.67 (1.3) &  9.43 (0.02) & 18.15 (0.1) & 3.5 (0.01) & 14.15 (0.08) & 16.5 (0.1) & 17.29 (0.09)   \tn
SiO$_2$  & 45.56 (3.9) & 46.8 (0.03) & 43.51 (0.02) & 49.3 (0.04) & 45.8 (0.02) & 43.8 (0.03) & 44.06 (0.02)   \tn
Ni (ppm) &  360 (93.6) &   222.7 (140.1) & 617.5 (674.5) & 114.0 (27.3) & 499.3 (543.8) & 884.3 (685.1) & 521.6 (554.1)   \tn
Co (ppm)  & 96 (44.2) &  33.0 (20.0) & 89.6 (81.6) & 16.8 (7.2) & 73.2 (66.4) & 100.6 (75.8) & 76.2 (71.9)   \tn
Nb (ppb) & 640 (0.1)  &  415.5 (13.2) & 655.7 (0.1) & 188.1 (20.2) & 687.8 (0.5) & 589.5 (2.3) & 658.8 (0.8)   \tn
V (ppm)  & 123 (12.3) & 89.1 (15.1) & 114.4 (0.9) & 71.9 (26.7) & 119.2 (2.2) & 107.0 (4.3) & 115.2 (1.4)   \tn
Cr (ppm)  & 6000 (420) &  3206.8 (258.4) & 4673.0 (95.0) & 2064.3 (412.9) & 4698.4 (159.7) & 4267.1 (131.7) & 4679.5 (107.6)   \tn
\tn
$\chi^2_{red}$ & Fit to both &  30.67 & 20.21 & 93.93 & 11.02 & 30.96 & 15.78 \tn

\bottomrule
\end{tabular*}
\begin{tablenotes}
      \item[a]{\citet{palme2014cosmochemical}}
      \item[b]{\citet{munker2003evolution}}
      \item[c]{\citet{yoshizaki2020composition}}
    \end{tablenotes}
\end{threeparttable}
\end{table*}

\subsection{Matching the mantle compositions of Earth and Mars}
The mantle compositions of Earth and Mars must have been established together from the same disk, but our model is unable to consistently reproduce their joint formation with a single set of redox conditions in the disk. 
When the model is fit jointly to the mantle composition of Earth and Mars, the calculated mantle composition of Earth consistently reproduces reality but the calculated mantle composition of Mars is typically significantly far from the estimated composition, as shown in Tables \ref{tblbothfittaylor} and \ref{tblbothfityoshizaki} assuming a Mars composition from either \citet{taylor2013bulk} or \citet{yoshizaki2020composition}.
When the model is fit solely to Earth it obtains a good match for an Earth-like planet but results generally in a poor approximation of Mars, as shown in Table \ref{tblearthfitwith5gpaswitch}. 
Similarly, when the model is fit solely to Mars it successfully produces a Mars-like planet, but fails to produce an accurate Earth-like planet, as shown in Tables \ref{tblmarsfittaylor} and \ref{tblmarsfityoshizaki} for a Mars with a \citet{taylor2013bulk} or \citet{yoshizaki2020composition} composition, respectively.
Regardless of which Martian mantle composition we used, our model produced a worse match to Earth and Mars' compositions (indicated by higher reduced chi-squared values) when attempting to simultaneously fit Earth and Mars than it did when fitting solely to Earth or Mars (see Table \ref{tblparam}).
Importantly, when fitting solely to Mars the model matches the Martian bulk and trace element abundances much better and more consistently than when the model is fit to reproduce Earth and Mars simultaneously.

\begin{table*}[width=\textwidth,cols=9,pos=t]
\caption{Results of fitting the model to the Earth mantle composition alone. Mantle abundances are shown in wt\% wt\% (or ppm/ppb where indicated) for the Earth-like planet for all simulations. $\chi^2_{red}$ values are low, indicating the simulated Earth mantles are close to observed values. This is in agreement with results in \citet{rubie2015accretion}.}\label{tblearthfitwith5gpaswitch}
\begin{threeparttable}
\begin{tabular*}{\textwidth}{@{}P{\eighttabcol}P{\eighttabcol}P{\eighttabcol}P{\eighttabcol}P{\eighttabcol}P{\eighttabcol}P{\eighttabcol}P{\eighttabcol}@{}}
\toprule
& Simulations: & 4:1-0.25-7  & 4:1-0.5-8  & 8:1-0.25-2  & 8:1-0.8-18 & i-4:1-0.8-4 & i-4:1-0.8-6 \tn 
\midrule
\underline{Earth} & Measured\tnote{a} \tn
FeO  & 8.1 (1.29) &   8.09 (0.02) & 8.09 (0.03) & 8.1 (0.02) & 8.09 (0.06) & 8.09 (0.04) & 8.09 (0.04)  \tn
SiO$_2$  & 45.4 (2.14) &  45.35 (0.1) & 45.63 (0.09) & 45.23 (0.09) & 45.29 (0.09) & 45.26 (0.1) & 45.4 (0.09)  \tn
Ni (ppm) &  1860 (5) &1793.8 (986.7) & 1823.5 (980.5) & 1751.0 (832.5) & 1649.1 (977.7) & 1749.8 (967.2) & 1746.4 (950.0)  \tn
Co (ppm)  & 102 (5) &   108.4 (55.6) & 111.4 (60.8) & 104.1 (48.5) & 115.2 (59.1) & 108.8 (58.8) & 111.7 (60.3)  \tn
Nb (ppb) &  595 (20) &   543.4 (16.7) & 560.3 (28.9) & 586.1 (20.9) & 583.0 (32.4) & 574.2 (21.5) & 569.2 (21.1)  \tn
Ta (ppb) & 43 (5) &  39.1 (0.8) & 40.8 (0.5) & 41.1 (0.4) & 41.0 (0.5) & 40.6 (0.6) & 40.7 (0.5)  \tn
Nb/Ta & 14.0 (0.3) \tnote{b} & 13.9 (0.7) & 14.0 (0.9) & 14.3 (0.6) & 14.2 (1.0) & 14.3 (0.7) & 14.2 (0.7) \tn
V (ppm)  & 86 (5) &  89.6 (15.8) & 88.1 (19.1) & 85.7 (21.1) & 84.0 (21.2) & 85.1 (19.5) & 85.9 (20.6)  \tn
Cr (ppm)  & 2520 (10) & 3099.2 (293.3) & 2930.5 (324.2) & 2756.2 (341.7) & 2703.0 (349.1) & 2823.6 (321.4) & 2777.1 (338.0)  \tn

\tn 

\underline{Mars} & Measured\tnote{c,d} \tn
FeO  & 14.67, 18.14 &  7.19 (0.02) & 22.77 (0.15) & 3.23 (0.005) & 27.12 (0.27) & 20.57 (0.14) & 23.4 (0.15)  \tn 
SiO$_2$  & 45.56, 43.85 &  46.74 (0.03) & 41.01 (0.02) & 49.12 (0.04) & 38.71 (0.03) & 41.74 (0.03) & 40.75 (0.02)  \tn
Ni (ppm) &  360, 330 &  170.8 (93.8) & 924.3 (1005.1) & 108.6 (22.3) & 1829.0 (1757.7) & 1178.8 (988.4) & 862.0 (938.8)  \tn
Co (ppm)  & 96, 71 & 24.6 (14.6) & 133.2 (109.3) & 14.3 (6.0) & 232.3 (145.6) & 141.4 (99.7) & 128.2 (106.8)  \tn
Nb (ppb) & 640, 501  &  404.2 (9.0) & 618.4 (0.2) & 174.3 (18.5) & 583.9 (0.3) & 562.7 (2.7) & 614.4 (0.1)  \tn
V (ppm)  & 123, 105 &  85.3 (15.0) & 108.2 (0.4) & 69.3 (26.8) & 102.3 (0.2) & 102.6 (3.9) & 107.5 (0.4)  \tn
Cr (ppm)  & 6000, 4990  &  3067.4 (269.3) & 4525.1 (53.9) & 1998.8 (404.7) & 4335.9 (29.9) & 4158.5 (99.0) & 4512.0 (50.6)  \tn
\tn
$\chi^2_{red}$ & Fit to Earth & 2.93 & 2.23 & 1.03 & 3.53 & 1.59 & 1.84 \tn
\bottomrule
\end{tabular*}
\begin{tablenotes}
      \item[a]{\citet{palme2014cosmochemical}}
      \item[b]{\citet{munker2003evolution}}
      \item[c]{\citet{yoshizaki2020composition}}
      \item[d]{\citet{taylor2013bulk}}
    \end{tablenotes}
\end{threeparttable}
\end{table*}

\begin{table*}[width=\textwidth,cols=9,pos=t]
\caption{Results of fitting the model to the Mars mantle composition \citep{taylor2013bulk} alone. Mantle abundances are shown in wt\% (or ppm/ppb where indicated) for Earth and Mars-like planets for all simulations. $\chi^2_{red}$ values are low, indicating the simulated Martian mantles are close to observed values.}\label{tblmarsfittaylor}
\begin{threeparttable}
\begin{tabular*}{\textwidth}{@{}P{\eighttabcol}P{\eighttabcol}P{\eighttabcol}P{\eighttabcol}P{\eighttabcol}P{\eighttabcol}P{\eighttabcol}P{\eighttabcol}@{}}
\toprule
& Simulations: & 4:1-0.25-7  & 4:1-0.5-8  & 8:1-0.25-2  & 8:1-0.8-18 & i-4:1-0.8-4 & i-4:1-0.8-6 \tn 
\midrule 
\underline{Earth} & Measured\tnote{a} \tn
FeO  & 8.1 (1.29) &  15.91 (0.06) & 2.07 (0.01) & 18.15 (0.06) & 8.41 (0.06) & 7.66 (0.01) & 4.96 (0.03)   \tn
SiO$_2$  & 45.4 (2.14) &  44.27 (0.07) & 16.25 (0.0) & 38.84 (0.05) & 40.68 (0.02) & 48.82 (0.03) & 42.84 (0.03)   \tn
Ni (ppm) &  1860 (5) &  1365.1 (870.4) & 190.6 (51.8) & 1208.4 (727.5) & 895.8 (387.0) & 229.5 (15.4) & 546.8 (242.7)   \tn
Co (ppm)  & 102 (5) &  118.6 (64.2) & 28.0 (12.1) & 137.8 (65.0) & 120.5 (26.7) & 25.4 (9.2) & 59.8 (22.6)   \tn
Nb (ppb) &  595 (20) &  574.4 (10.1) & 228.1 (1.0) & 518.0 (11.6) & 180.5 (6.9) & 500.3 (9.8) & 260.7 (25.8)   \tn
Ta (ppb) & 43 (5) &  36.0 (0.1) & 12.8 (0.2) & 32.6 (1.0) & 28.5 (1.9) & 38.7 (0.2) & 33.8 (2.2)   \tn
Nb/Ta & 14.0 (0.3)\tnote{b} & 15.9 (0.3) & 19.0 (0.4) & 16.2 (0.9) & 6.4 (0.7) & 13.2 (0.3) & 7.9 (1.0) \tn
V (ppm)  & 86 (5) & 101.3 (8.1) & 40.4 (1.5) & 83.8 (7.7) & 50.7 (18.3) & 105.2 (13.0) & 51.8 (21.3)   \tn
Cr (ppm)  & 2520 (10) & 3876.0 (172.4) & 1665.1 (55.4) & 3337.7 (125.8) & 1835.1 (226.0) & 3697.8 (309.0) & 1761.0 (271.7)   \tn

\tn
\underline{Mars} & Measured\tnote{c} \tn
FeO  & 18.14 (0.5) &  18.15 (0.04) & 18.2 (0.03) & 18.16 (0.04) & 18.14 (0.03) & 18.13 (0.01) & 18.24 (0.03)   \tn
SiO$_2$  & 43.85 (1.0) &  43.57 (0.02) & 43.39 (0.01) & 43.64 (0.01) & 43.36 (0.01) & 43.18 (0.01) & 43.31 (0.01)   \tn
Ni (ppm) &  330 (54.5) &  333.9 (242.6) & 324.1 (175.7) & 330.4 (203.4) & 325.9 (184.8) & 379.8 (48.5) & 339.5 (179.1)   \tn
Co (ppm)  & 71 (12.5) &  66.5 (40.0) & 71.4 (34.9) & 74.1 (37.8) & 69.2 (29.9) & 62.5 (27.0) & 56.5 (29.5)   \tn
Nb (ppb) &  501 (0.035) &  592.0 (3.6) & 645.7 (0.6) & 639.3 (0.7) & 624.2 (0.2) & 587.0 (2.3) & 619.4 (0.3)   \tn
V (ppm)  & 105 (5.25) &  107.4 (4.3) & 112.6 (1.6) & 112.0 (2.4) & 110.5 (1.7) & 107.7 (3.5) & 110.1 (1.9)   \tn
Cr (ppm)  & 4990 (210) &  4296.7 (125.4) & 4544.1 (116.5) & 4478.2 (130.7) & 4572.1 (84.4) & 4376.1 (105.7) & 4553.3 (84.0)   \tn
\tn
$\chi^2_{red}$ & Fit to Mars &  1.52 & 2.37 & 2.2 & 1.63 & 2.01 & 2.22 \tn
\bottomrule
\end{tabular*}
\begin{tablenotes}
      \item[a]{\citet{palme2014cosmochemical}}
      \item[b]{\citet{munker2003evolution}}
      \item[c]{\citet{taylor2013bulk}}
    \end{tablenotes}
\end{threeparttable}
\end{table*}

\begin{table*}[width=\textwidth,cols=9,pos=t]
\caption{Results of fitting the model to the Mars mantle composition \citep{yoshizaki2020composition} alone. Mantle abundances are shown in wt\% (or ppm/ppb where indicated) for Earth and Mars-like planets for all simulations. $\chi^2_{red}$ values are low, indicating the simulated Martian mantles are close to observed values.}\label{tblmarsfityoshizaki}
\begin{threeparttable}
\begin{tabular*}{\textwidth}{@{}P{\eighttabcol}P{\eighttabcol}P{\eighttabcol}P{\eighttabcol}P{\eighttabcol}P{\eighttabcol}P{\eighttabcol}P{\eighttabcol}@{}}
\toprule
& Simulations: & 4:1-0.25-7  & 4:1-0.5-8  & 8:1-0.25-2  & 8:1-0.8-18 & i-4:1-0.8-4 & i-4:1-0.8-6 \tn 
\midrule 
\underline{Earth} & Measured\tnote{a} \tn
FeO  & 8.1  (1.29) &   14.66 (0.06) & 12.64 (0.06) & 21.47 (0.06) & 9.69 (0.09) & 11.05 (0.01) & 9.5 (0.04)   \tn
SiO$_2$  & 45.4 (2.14) &   45.17 (0.07) & 46.04 (0.07) & 41.28 (0.03) & 47.78 (0.06) & 47.42 (0.01) & 47.18 (0.08)   \tn
Ni (ppm) &  1860 (5) & 1437.3 (877.3) & 1252.2 (880.8) & 1066.9 (462.2) & 1156.1 (735.9) & 227.1 (20.7) & 1418.8 (856.6)   \tn
Co (ppm)  & 102 (5) &  140.6 (56.6) & 97.4 (70.6) & 250.8 (42.0) & 117.2 (44.7) & 28.9 (12.0) & 94.9 (58.3)   \tn
Nb (ppb) &  595 (20) &  666.8 (1.7) & 713.3 (0.4) & 468.3 (2.9) & 639.7 (19.0) & 712.8 (0.3) & 738.1 (1.7)   \tn
Ta (ppb) & 43 (5) & 36.4 (0.04) & 37.9 (0.01) & 32.4 (0.2) & 38.8 (0.04) & 37.9 (0.01) & 39.5 (0.02)   \tn
Nb/Ta & 14.0  (0.3)\tnote{b} & 18.5 (0.07) & 19.3 (0.02) & 14.6 (0.2) & 16.8 (0.5) & 19.2 (0.01) & 18.9 (0.05)\tn
V (ppm)  & 86 (5) &  112.2 (5.3) & 118.6 (3.6) & 91.9 (7.2) & 107.6 (14.6) & 124.2 (1.5) & 119.1 (6.8)   \tn
Cr (ppm)  & 2520 (10) &  4254.6 (197.3) & 4605.2 (171.5) & 3633.6 (135.6) & 3489.0 (377.4) & 4971.6 (154.6) & 4346.0 (264.2)   \tn

\tn
\underline{Mars} & Measured\tnote{c} \tn
FeO  & 14.67 (1.3) &   14.77 (0.03) & 14.66 (0.06) & 14.76 (0.02) & 14.65 (0.06) & 14.64 (0.01) & 14.68 (0.06)   \tn
SiO$_2$  & 45.56 (3.9) &45.58 (0.02) & 45.41 (0.02) & 45.71 (0.02) & 45.56 (0.02) & 45.16 (0.01) & 45.48 (0.02)   \tn
Ni (ppm) &  360 (93.6) &  361.7 (231.9) & 375.3 (394.9) & 357.3 (142.2) & 375.5 (383.8) & 372.9 (35.9) & 378.2 (380.5)   \tn
Co (ppm)  & 96 (44.2) &  72.3 (31.0) & 55.4 (55.4) & 85.9 (20.8) & 64.0 (52.2) & 46.6 (18.5) & 54.5 (54.0)   \tn
Nb (ppb) & 640 (0.1)  & 685.3 (0.3) & 684.0 (0.05) & 686.4 (0.3) & 685.9 (0.1) & 680.3 (0.1) & 685.0 (0.05)   \tn
V (ppm)  & 123 (12.3) &  118.9 (2.1) & 119.2 (1.2) & 119.9 (1.3) & 119.5 (1.3) & 118.9 (0.9) & 119.4 (1.2)   \tn
Cr (ppm)  & 6000 (420) &  4692.6 (160.4) & 4799.8 (126.5) & 4842.9 (132.0) & 4802.6 (134.0) & 4849.0 (113.7) & 4820.6 (124.7)   \tn
\tn
$\chi^2_{red}$ & Fit to Mars & 1.09 &  1.24 & 0.77 & 1.07 & 1.4 & 1.23 \tn
\bottomrule
\end{tabular*}
\begin{tablenotes}
      \item[a]{\citet{palme2014cosmochemical}}
      \item[b]{\citet{munker2003evolution}}
      \item[c]{\citet{yoshizaki2020composition}}
    \end{tablenotes}
\end{threeparttable}
\end{table*}

The parameters that control the steepness and shape(s) of the Fe and Si oxidation gradients as well as the equilibration pressure factors in Table \ref{tblparam} are significantly different between the models that reproduce Earth well and those that reproduce Mars well.
The pressure factors are generally lower for models that best produce Mars than those that best produce Earth. 
Such a result is consistent with Mars forming as a stranded embryo. 

\begin{figure}
    \centering
    \includegraphics[width=\columnwidth]{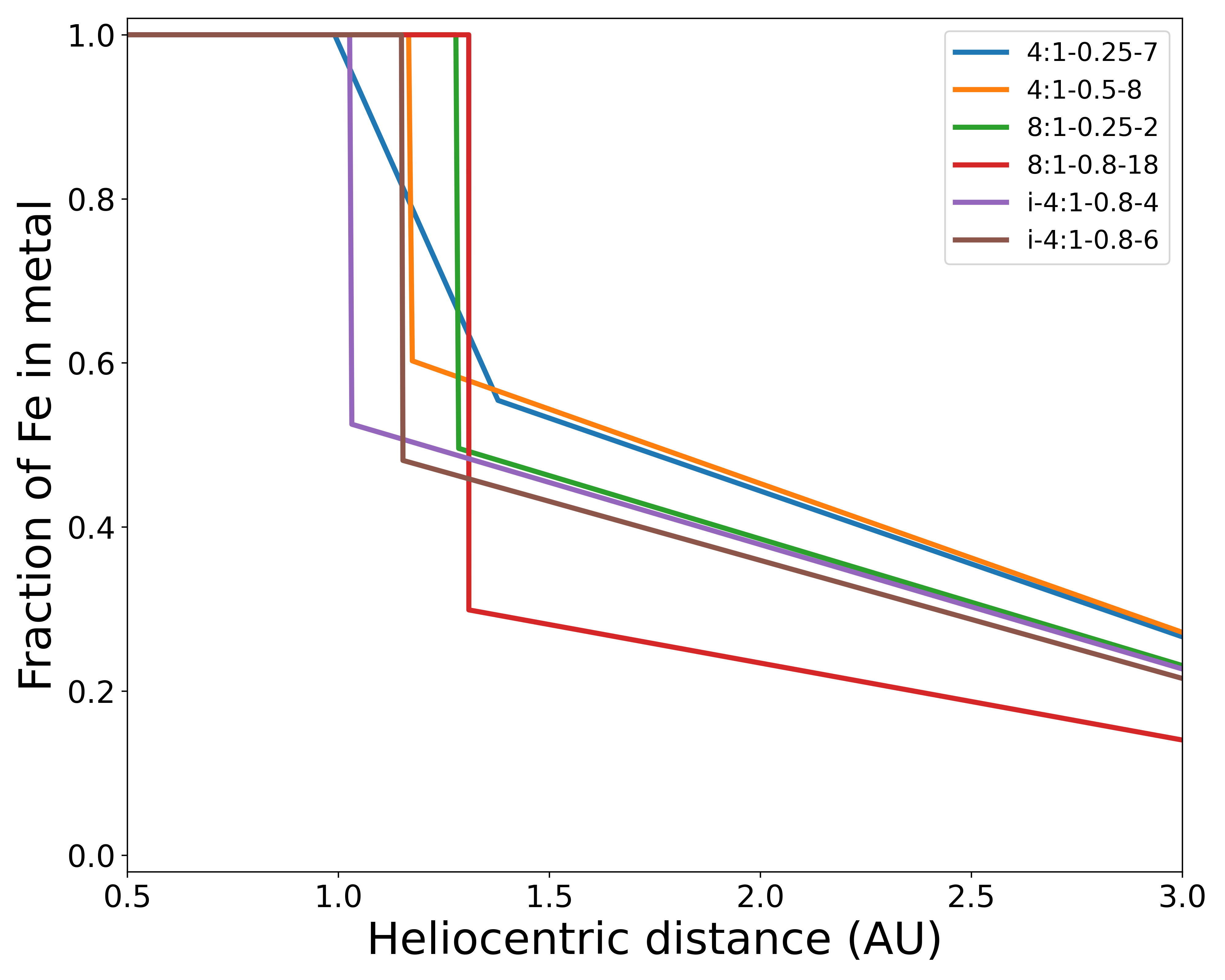}
    \caption{Earth best fit oxidation gradients for simulations 1 - 6, with each color corresponding to the simulation shown in the legend. 
    See Table \ref{tbl1} for the properties of each simulation.
    Each line represents the best-fit initial conditions to produce a compositionally representative Earth for a different Grand Tack simulation.
    The horizontal axis shows location in the protoplanetary disk, in astronomical units.
    The vertical axis shows the fraction of iron in metal in the planetesimals and embryos that comprise the initial disk.
    There is a high degree of similarity in the best fit oxidation gradient that produces Earth in different simulations.
    \label{fig:earth_fit_disk}}
\end{figure}

The iron oxidation gradient for each simulation controls the initial Fe redox state of the starting embryos and planetesimals in the protoplanetary disk, and it is determined by the parameters: $\delta$(1), $\delta$(2), and $X^{met}_{Fe}$(2). 
Uncertainty about the best fit value in these parameters indicates a large range in potential iron oxidation gradients that may properly reproduce the Martian mantle composition.
We note that there is a difference between the Earth-producing oxidation gradients in this study and those presented in \citet{rubie2015accretion}.
Here, as shown in Figure \ref{fig:earth_fit_disk}, Earth-producing disks display steep oxidation gradients, in contrast to the more gradual oxidation gradients found by \citet{rubie2015accretion}. 
In \citet{rubie2015accretion}, a gradual gradient was shown to reproduce an Earth that accretes from increasingly oxidized material, as suggested in \citet{wade2005core}. 
This type of gradient and subsequent heterogeneous accretion is consistent with an evolving water-line in the protoplanetary disk \citep{cuzzi2004material, monteux2018water} as well as evidence from silver isotope data \citep{schonbachler2010heterogeneous}.
As a result of the more rigorous exploration of parameter space in this study, we have found a global minimum parameter set wherein there is a near-degeneracy between parameters $\delta$(1) and $\delta$(2). 
This indicates that the best fit oxidation gradient for an Earth-producing disk may indeed resemble something closer to a step function than a smooth gradient. 
Since interaction of embryos and planetesimals with water controls the oxidation gradient in the disk, the resemblance of the oxidation gradient to a step-function may suggest that oxidation in the protoplanetary disk was set by a single snapshot of the location of the ice line at the time of the majority of planetary accretion, contrary to prior studies.

\begin{figure}
    \centering
    \includegraphics[width=\columnwidth]{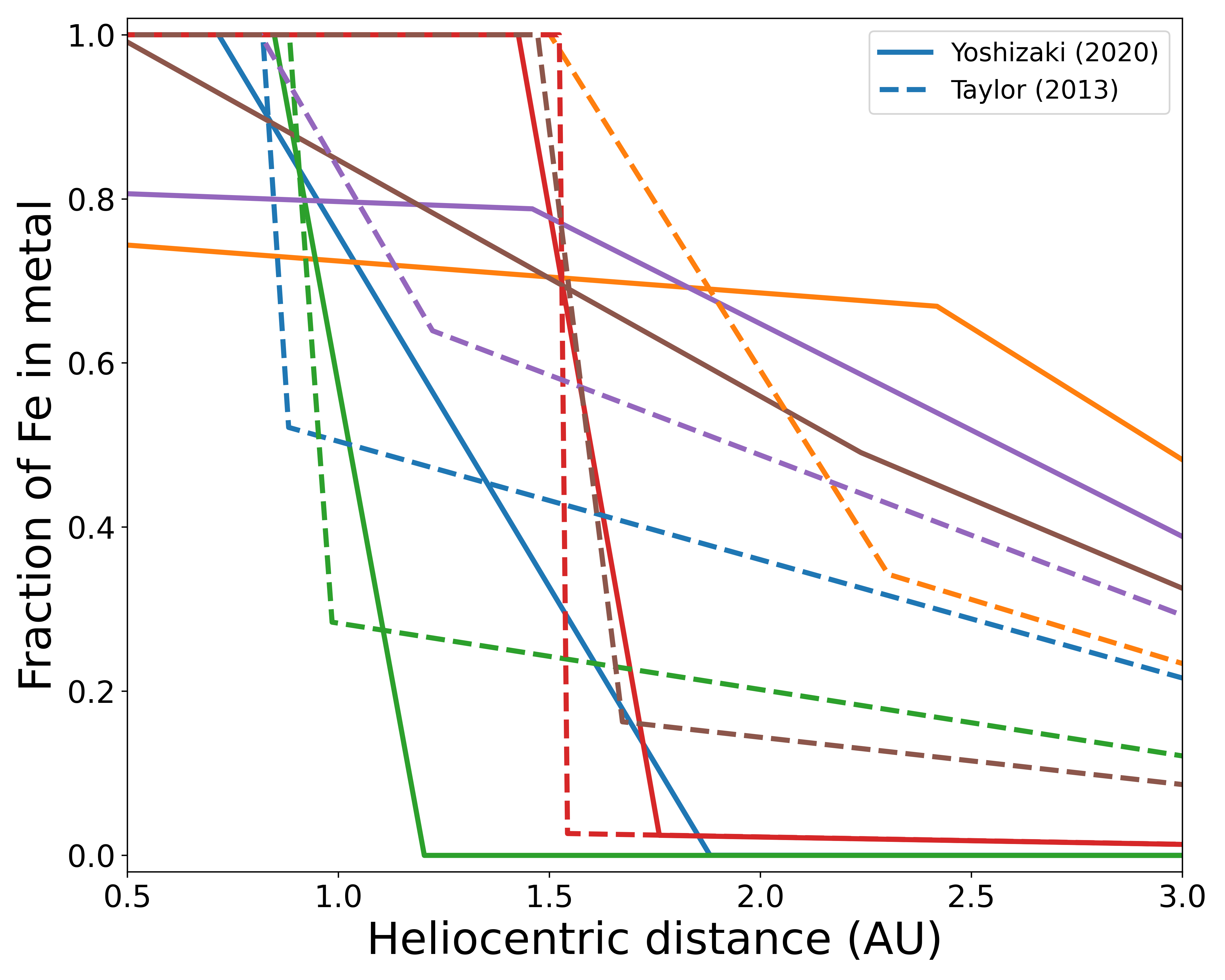}
    \caption{Mars best fit oxidation gradients for simulations 1 - 6.
    Colors correspond to each simulation as shown in Figure \ref{fig:earth_fit_disk}, see Table \ref{tbl1} for the properties of each simulation.
    Each line represents the best fit initial conditions of the disk to produce a compositionally accurate Mars for a different Grand Tack simulation.
    Solid lines show the oxidation gradients when Mars' mantle is assigned the composition of \citet{yoshizaki2020composition}, and dashed lines show gradients for when Mars' mantle is assigned the composition of \citet{taylor2013bulk}.
    The horizontal axis shows location in the protoplanetary disk, in astronomical units.
    The vertical axis shows the fraction of iron in metal in the planetesimals and embryos that comprise the initial disk.
    There is significant variability in the best fit oxidation gradient to produce a Mars analog due to stochastic formation processes becoming exaggeratedly varied as a result of Mars' small mass and rapid growth.
    \label{fig:mars_fit_disk}}
\end{figure}

Comparison between the Earth-producing oxidation gradients shown in Figure \ref{fig:earth_fit_disk} and the Mars-producing oxidation gradients shown in Figure \ref{fig:mars_fit_disk} reveals their notable differences. 
The Earth-producing oxidation gradients are nearly the same regardless of which simulation is used, as shown in Figure \ref{fig:earth_fit_disk} with parameters listed in Table \ref{tblparam}.
Each simulation's best fit disk becomes rapidly oxidized near 1.3 AU and then have nearly no iron in metallic form exterior to that point. 
The best Mars-producing oxidation gradients, however, are all different from one another, as shown in Figure \ref{fig:mars_fit_disk}.
Another way to express these differences is to look at the particularly large standard deviations of the best-fit values for $\delta$(1) and $\delta$(2) in Table \ref{tblparam} for scenarios where the model attempts to recreate the Martian mantle composition.
While the Mars-producing oxidation gradients all become more oxidized at greater heliocentric distances, the rate at which this change occurs has no pattern and is completely incompatible with the gradients that produce Earth.
Comparing the parameters that describe the best fit oxidation gradients for Earth and Mars, if the uncertainties about the mean best-fit values overlap, e.g. for $\delta$(1) and $\delta$(2), it is almost always due to the much larger uncertainty about the Mars best fit.
In fact, the uncertainty around the mean best-fit Earth values rarely encompasses the mean best-fit values for Mars.
Furthermore, the mean and standard deviations in parameter values that reproduce Earth are usually very similar across simulations, while the parameters for Mars vary wildly, due in large part to the variation in the Mars embryo's origin and accretion history.

In general, the simultaneous fit cannot reproduce the mantle composition of both planets well, either with respect to major (Fe, Si) or trace moderately siderophile elements (Ni, Co, Nb, Ta, V, and Cr).
The model often produces a better approximation of Earth's mantle abundances than it does for Mars' mantle--although each mantle composition was weighted equally as a constraint in the chi square fitting.
For Earth, the mantle concentrations of FeO and SiO$_2$ are close to observed values, while for Mars, the calculated FeO and SiO$_2$ abundances vary significantly from the Martian estimated mantle abundances, regardless of which Martian mantle composition (\citet{taylor2013bulk} or \citet{yoshizaki2020composition}) is used, as shown in Tables \ref{tblbothfittaylor} and \ref{tblbothfityoshizaki}. 
For example, simultaneous-fit Martian analog FeO varies between 3.5 and 18.15 wt \%, depending on the simulation. 
Martian values of Cr are always underestimated, but there is not a pattern to the abundances of the other elements; depending on the simulation, the other Martian elemental abundances are either under- or overestimated, regardless of which Martian composition is used.
In the joint fit, the calculated composition of the Martian mantle varies considerably between simulations unlike the simulated Earth mantle, this is a fundamental result reflecting the different growth histories and feeding zones of Earth and Mars.

\begin{figure}
    \centering
    \includegraphics[width=\columnwidth]{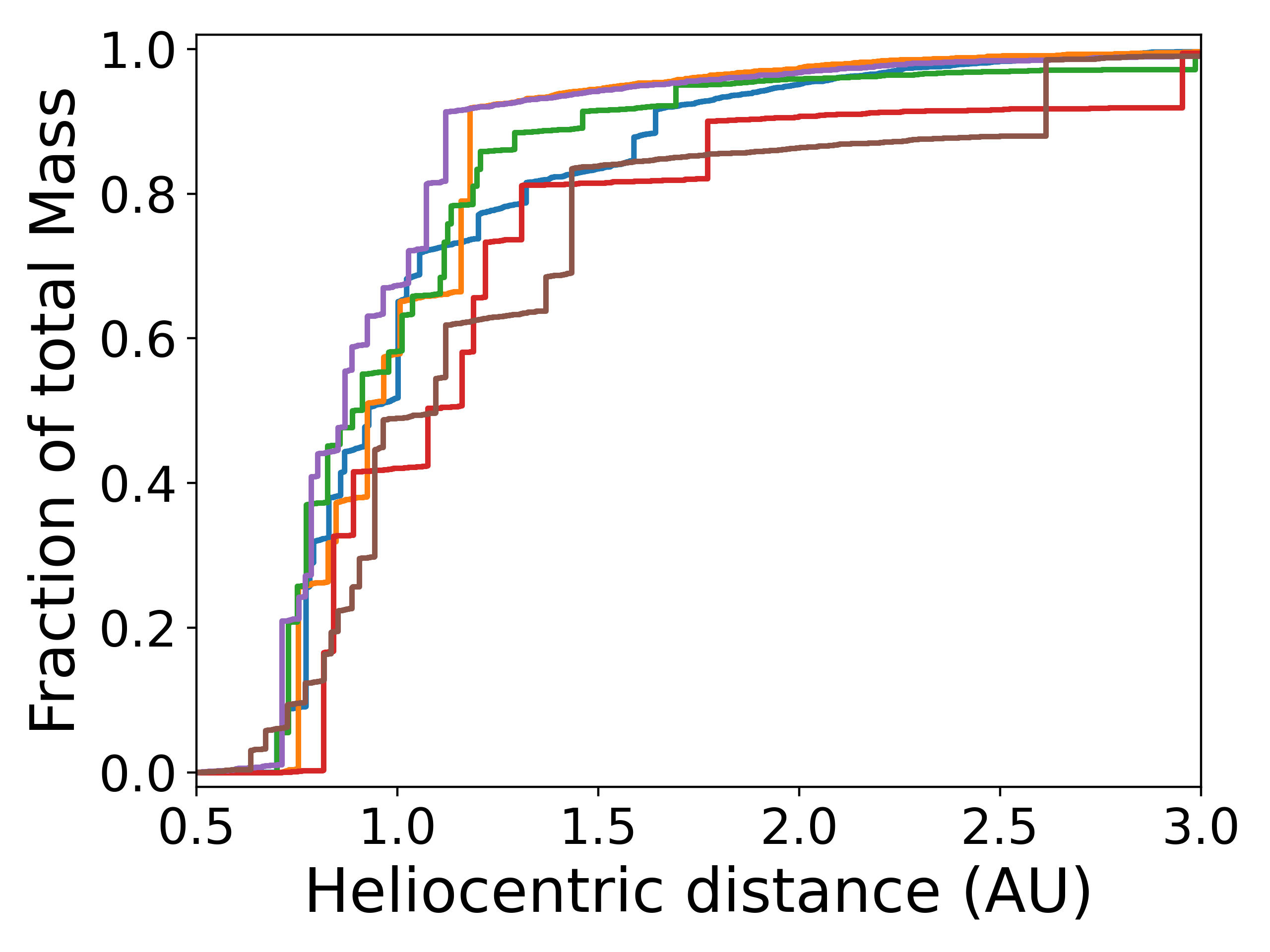}
    \caption{Source regions of the material that builds the Earth-like planet.
    Each curve represents the growth of an Earth-like planet in one of six different Grand Tack scenarios.
    Colors correspond to each simulation as shown in Figure \ref{fig:earth_fit_disk}, see Table \ref{tbl1} for the properties of each simulation.
    The horizontal axis shows where each planetesimal or embryo building block originally started in the simulation.
    The vertical axis shows cumulative amount of the mass in the Earth-like planet is built from planetesimals and embryos from within a given heliocentric distance.
    Each vertical increase shows the growth of the body in an impact.
    These are represented as step functions because growth occurs as the result of impact events that are modeled as 100 percent efficient (i.e. all of the projectile and target material are combined in the new embryo).
    \label{fig:earth_accumulation}}
\end{figure}

\subsection{Understanding how the feeding zones of Earth and Mars determine their mantle chemistry}

The reason for the difference between the best fit Earth-forming and Mars-forming iron oxidation gradients is that, in addition to having distinct bulk chemistries, the accretion histories of Earth and Mars are very different from one another.
As shown in Figure \ref{fig:earth_accumulation}, the Earth-like planet is built from material originating from the whole inner Solar System; it is built mainly of material from 0.5 to 2.5 AU, with most of the accumulated mass originating around 1 AU and a small mass of distant material originating beyond 4.5 AU. 
Roughly 80 percent of the Earth-like planets' mass originates between 0.5 AU and 1.5 AU, but within this boundary no particular location  contributes preferentially over another to the final mass of Earth.
This wide averaging of the disk explains why the oxidation gradient necessary to reproduce Earth is very similar regardless of which of the 6 simulations is in question, as shown in Figure \ref{fig:earth_fit_disk}.
Additionally, since each Earth-like planet contains roughly half of the mass of the inner-planet forming disk, Earth samples much of the entire inner disk.

\begin{figure}
    \centering
    \includegraphics[width=\columnwidth]{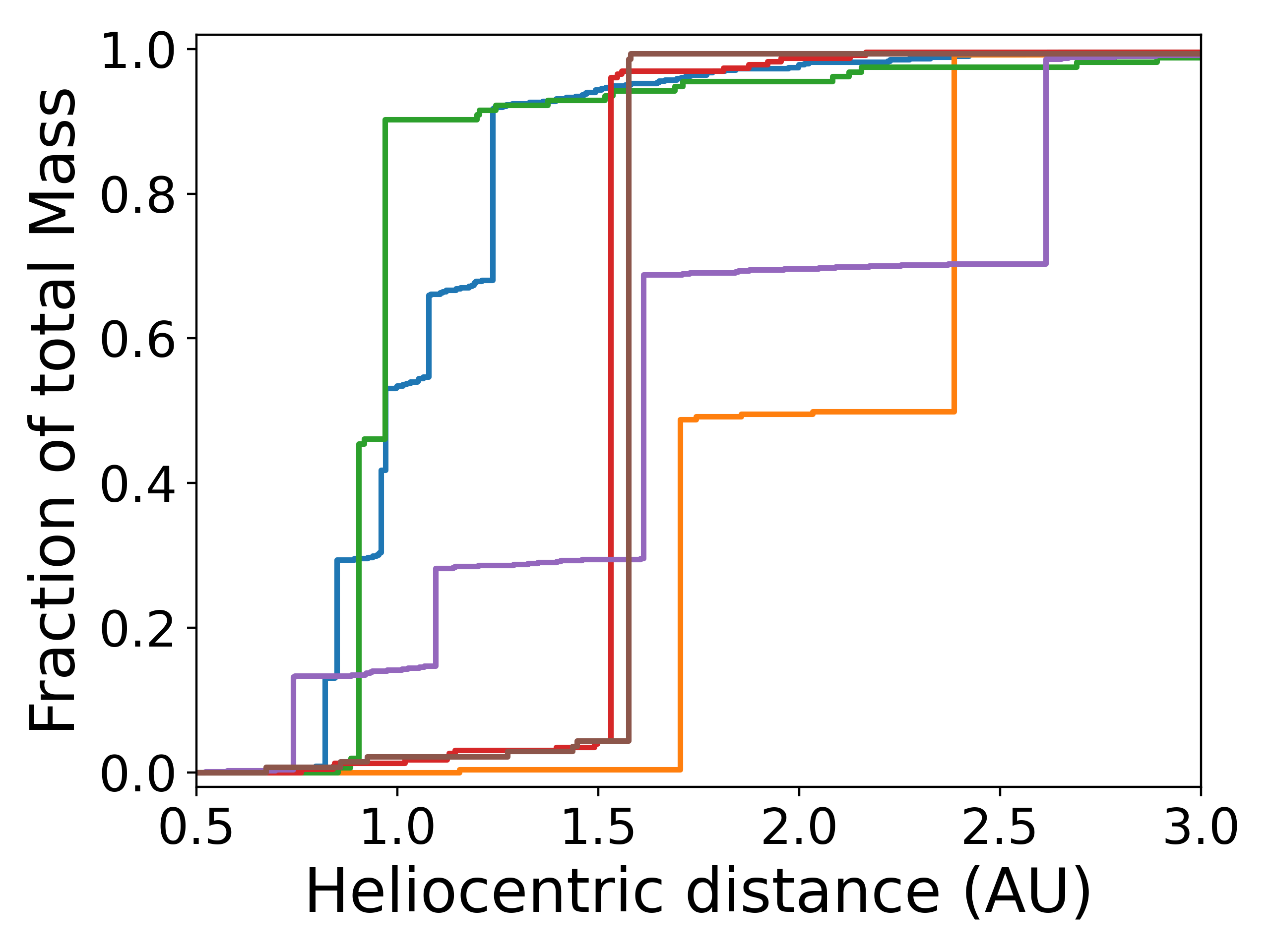}
    \caption{Source regions of the material that builds the Mars-like planet.
    Each curve represents the growth of a Mars-like planet in one of six different Grand Tack scenarios.
    Colors correspond to each simulation as shown in Figure \ref{fig:earth_fit_disk}, see Table \ref{tbl1} for the properties of each simulation.
    The horizontal axis shows where each planetesimal or embryo building block originally started in the simulation.
    The vertical axis shows how much cumulative mass in the final Mars-like planet is contributed by planetesimals and embryos from within a given heliocentric distance.
    The vertical steps represent an impact and corresponding episode of planet growth with an embryo originating at a certain location.
    \label{fig:mars_accumulation}}
\end{figure}

In contrast, the feeding zone of the Mars-like planets have great variability, as shown in Figure \ref{fig:mars_accumulation}. 
The large jumps in the shown cumulative distribution functions demonstrate that each Mars-like planet is built from only a few distinct source regions, where embryos initially grew out of annuli of local material.
A single locally-sourced embryo may contribute a majority of the Mars-like planet's mass.
Indeed, often a single embryo contributes greater than 80 percent of Mars' final mass. 
The Mars-like planet's ultimate composition is dependent on a far fewer number of individual bodies than the Earth-like planet's, making it more influenced by stochastic differences between simulations.
The few bodies that accrete to form Mars can come from multiple different source regions in the disk due to radial mixing in the disk.
The lower number of building blocks that are needed to create Mars prohibits the emergence of the type of averaging of source location that is seen in the growth of the Earth-like planet.

The much greater number of embryos that build the Earth-like planet means that no one embryo or location in the disk has a controlling influence on the final composition of the body. 
In contrast, the source of the Martian embryo (or couple of embryos) does not reflect the disk at large, but is instead assembled from locally distinct material from specific locations in the disk before being scattered to a location consistent with Mars' semi-major axis.
We expect this trend to hold broadly beyond just simulations from the Grand Tack scenario; since Mars represents a negligible (~$5\%$) mass of the terrestrial disk, it will always be the case that Mars is unlikely to represent the whole disk.
Indeed, Mars' small size means that not only does it not affect the dynamics of the inner disk but also that it can be sourced from a wide variety of locations within the disk and subsequently scattered to its final location.

The takeaway is that the differences between best fit oxidation gradients for Earth-producing and Mars-producing disks are driven by the stark differences in their feeding zones, which explains why a simultaneous fit of both planets' compositions is so difficult.
Earth grows from many embryos and impacts and its best fit formation disk is consistent between simulations, whereas Mars' growth is more stochastic and correspondingly does not have a consistent best fit disk reproduced by multiple simulations. 
Both the disks share the same basic features; they are increasingly oxidized at greater heliocentric distances, in agreement with results from \citet{rubie2015accretion}. 
Ultimately however, Mars' distinct chemistry and chaotic accretion mean that the oxidation gradients needed to produce Mars are incompatible with those that produce Earth.

\subsection{Reconciling the accretion histories of Earth and Mars}
\label{section:reconcile_model}

It is problematic that the best fit iron oxidation gradients the model finds to produce Earth are different than those it finds to produce Mars because, in reality, Earth and Mars must have formed from the same circumstellar disk with a single oxidation gradient.
Therefore, it is essential to reconcile the differences between the Earth and Mars forming oxidation gradients. 

In a strong radial mixing paradigm such as in the Grand Tack, the feeding zones of Earth and Mars are fundamentally different, as shown in Figures \ref{fig:earth_accumulation} and \ref{fig:mars_accumulation} and described in the text above.
Note that the Earth-like planets are formed from a consistent feeding zone, whereas the feeding zones of Mars-like planets are stochastic and scattered about the inner disk.
Therefore, this explains why the oxidation gradients that reproduce Earth's mantle composition are similar across each of the simulations, whereas the oxidation gradients that reproduce Mars' mantle composition are strongly varying.
Due to its overall small mass and rapid formation requiring construction from a single to a few locally-sourced embryos, the building blocks of Mars are not widely sourced from throughout the disk.
This is in strong contrast to Earth, which represents effectively half of the inner disk's mass and has a feeding zone that extends across the entire disk.
This explains why the oxidation gradients that reproduce the terrestrial mantle are consistent across simulations whereas those for the martian analogs are not.

We then ask, what changes to Mars' feeding zone must occur in order to reproduce the mantle of Mars in a disk that also reproduces the mantle of Earth?
To answer this question, we must determine the origin of Mars' building blocks within an Earth-forming disk, and this requires four steps.
First, we identify fitted disk oxidation models that reproduce an accurate terrestrial mantle composition in the Earth-like planets, as we have already done above.
Second, similarly, we identify fitted disk oxidation models that reproduce an accurate Martian mantle composition in the Mars-like planets, as we again have already done above.
Third, we examine the Mars-producing iron oxidation gradient and find the FeO content for each corresponding planetesimal and embryo that forms the Mars-like planet.
Fourth, we calculate where each of those Mars-forming planetesimals or embryos would need to originate in the Earth-producing disk (from step 1) in order for each to have the same FeO content (as in step 2). 
For example, if an individual planetesimal that builds Mars has a fraction of bulk Fe as metal of 0.6, we then find where in the Earth-producing disk's iron oxidation gradient the Fe metal fraction is 0.6 and determine that this is the ideal origin for the individual Mars component in question.

Manipulating the N-body history in this way is a reasonable strategy specifically because the goal of this study is not to determine a precise accretion history for Mars. 
This is true for any astrophysical N-body simulation of the forming terrestrial planets since the stochasticity due to dynamical chaos makes determining exact formation histories of planets impossible.
Rather, the goal here is to reconcile Martian mantle chemistry and potential accretion history with that of the Earth.
Fe and Si are the only elements that vary on the basis of heliocentric distance within our model, and the shape of the Si oxidation gradient (a step function controlled by $X^{met}_{Si}$ and $\delta$(1)) is much more similar between the Mars-fit model and the Earth-fit model than the Fe oxidation gradient.
The Mars-fit model produces Martian analogs with much more accurate bulk and trace element abundances than the Martian analogs produced by the Earth-fit or simultaneous Earth-and-Mars-fit models. 
Thus, in order to create the Martian mantle in an Earth-producing disk, we have proposed a different origin within the disk for Mars' building blocks, violating only the N-body accretion history of the Mars body, but leaving its mantle chemistry (with the exception of small differences in the redox state of Si) effectively unaltered.
By calculating alternate locations of the origin of Mars' building blocks within the disk based on their FeO content, we are able to identify broad trends that can constrain Mars' origin within the protoplanetary disk.

\begin{figure} 
    \centering
    \includegraphics[width=\columnwidth]{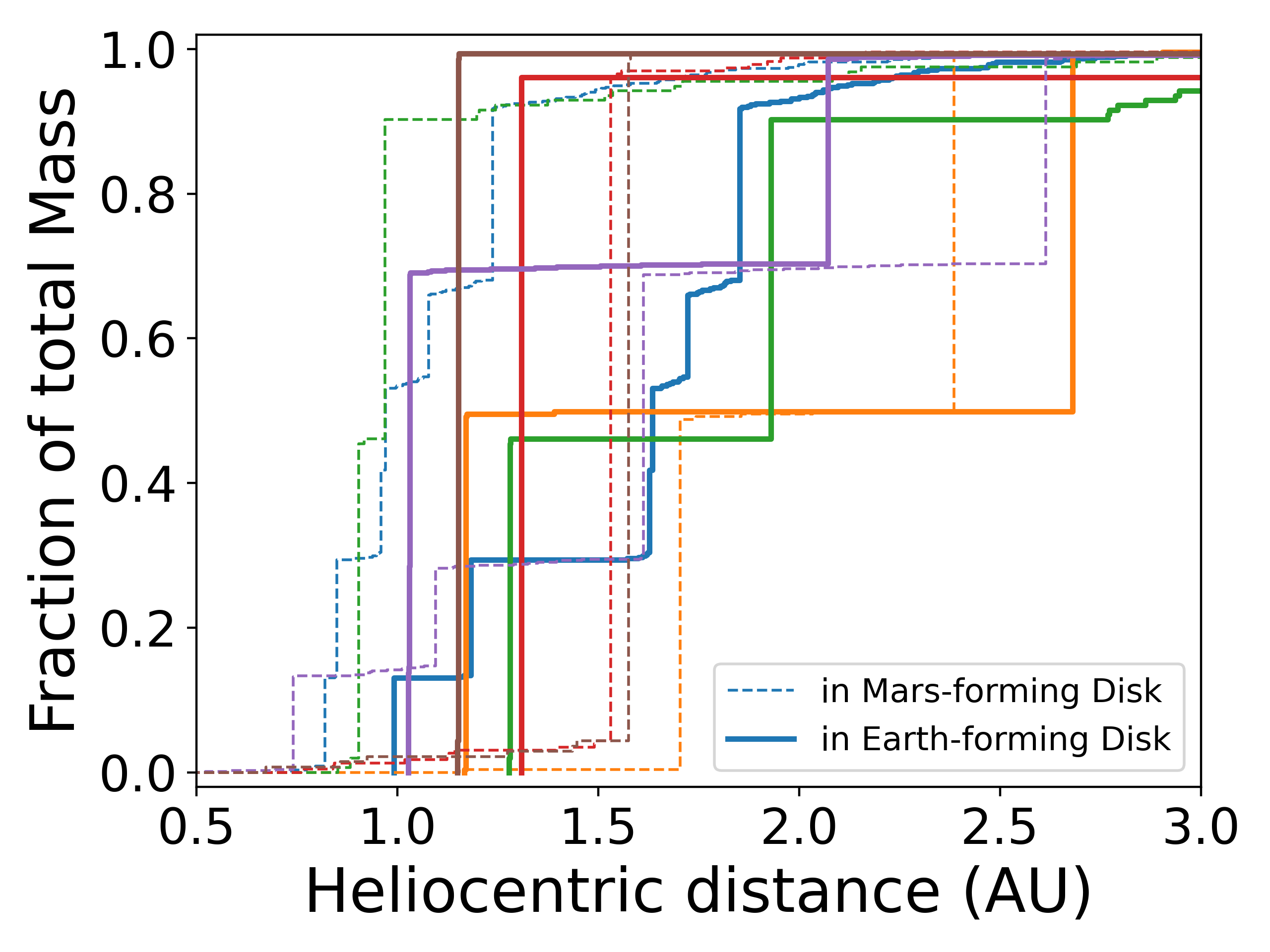} 
    \caption{The ideal mass accumulation for Mars embryos in the Earth-forming protoplanetary disk for \citet{taylor2013bulk} composition.
    Colors correspond to each simulation as in Figure \ref{fig:earth_fit_disk}, see Table \ref{tbl1} for the properties of each simulation.
    Dashed lines indicate unaltered growth patterns, as is shown in Figure \ref{fig:mars_accumulation}.
    When the same component masses are sourced from the best fit Earth forming disk, they must originate in a band of a region of the disk primarily between 1.0 and 2.0 AU with an appropriate oxidation state to build Mars, shown in solid lines.
    \label{fig:idealtaylor}}
\end{figure}

\begin{figure} 
    \centering
    \includegraphics[width=\columnwidth]{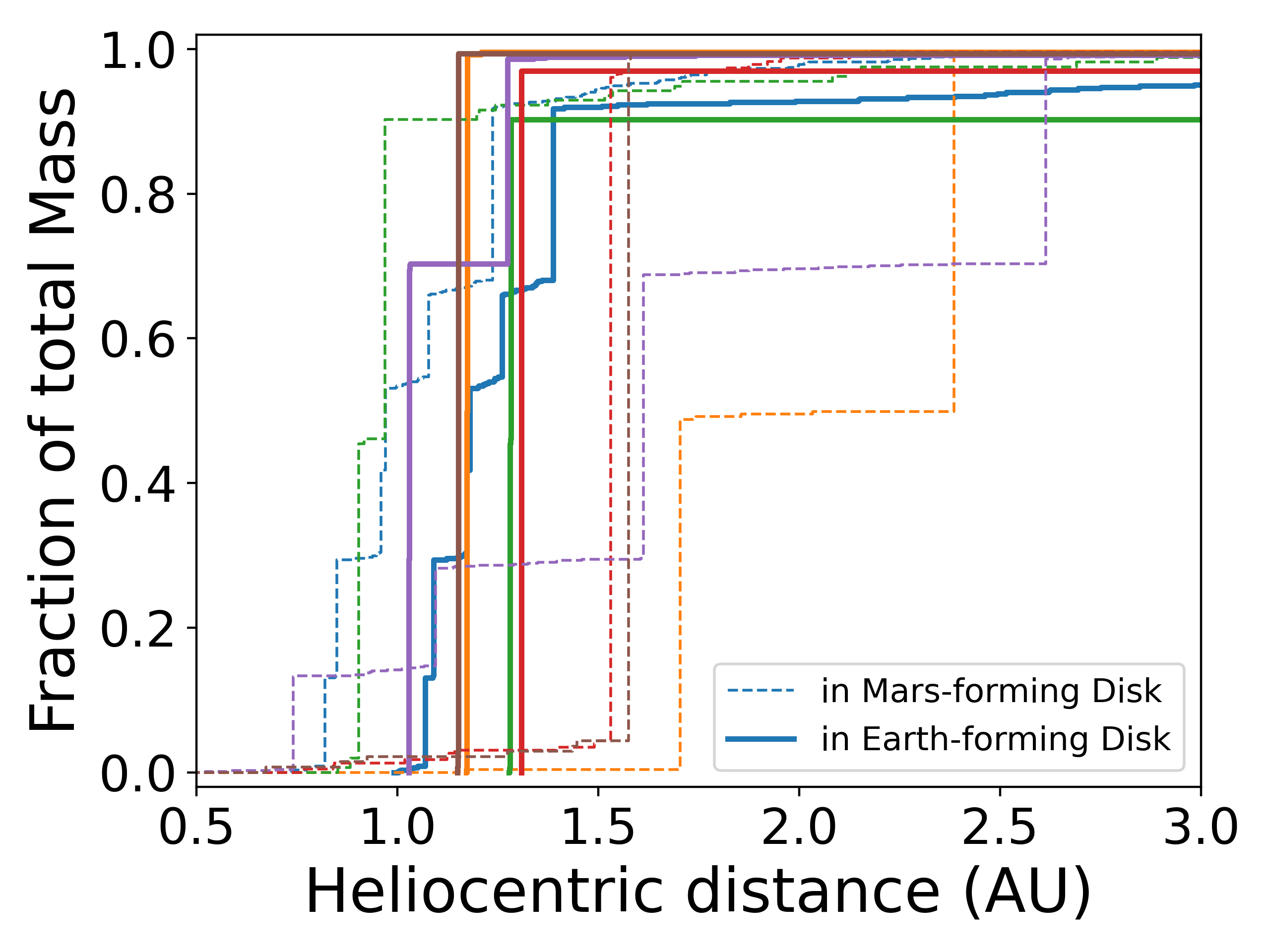} 
    \caption{The ideal mass accumulation for Mars embryos in the Earth-forming protoplanetary disk for \citet{yoshizaki2020composition} composition.
    Colors correspond to each simulation as in Figure \ref{fig:earth_fit_disk}, see Table \ref{tbl1} for the properties of each simulation.
    Dashed lines indicate unaltered growth patterns, as is shown in Figure \ref{fig:mars_accumulation}.
    When the same component masses are sourced from the ideal Earth forming disk, they must originate in a band of a more constrained region of the disk between 1.0 and 1.3 AU with an appropriate oxidation state to build Mars, shown in solid lines. 
    \label{fig:idealyoshi}}
\end{figure}

By completing these steps for each simulation, a trend is revealed: the Mars-like planet's constituent parts must come from a relatively constrained region of the outer more oxidized region of the protoplanetary disk that forms Earth.
The location that Mars' building blocks would have to originate within the Earth-producing disk are shown in Figures \ref{fig:idealtaylor} and \ref{fig:idealyoshi}.
Figure \ref{fig:idealtaylor} shows the change in location to reproduce the \citet{taylor2013bulk} Martian composition in an Earth forming disk, and Figure \ref{fig:idealyoshi} shows the change in location for a \citet{yoshizaki2020composition} Martian mantle composition.
The dotted lines show the unaltered mass accumulation history of the Martian component planetesimals and embryos, which is identical to that shown in Figure \ref{fig:mars_accumulation}. 
The solid lines show the locations at which these embryos and planetesimals ought to originate in the Earth-fit disk. 
The rest of the mass of the Mars body in each simulation comes from planetesimals, which are not shown in this table and are far more numerous, but exhibit a similar trend, clustering in the region between 1.0 and 2.0 AU.

Mars' building blocks must originate near or exterior to 1.0 AU because the iron oxidation gradient of the disk that best reproduces Earth is very reducing interior to that distance.
Specifically, the model parameter $\delta$(1) sets the end of the region of fully-reduced iron in the inner disk and in an Earth-producing disk its mean value is never less than 0.96 AU, with small standard deviation, regardless of the simulation (see Table \ref{tblparam}). 
Interior to this parameter all iron is fully reduced, and while Earth's composition requires both reduced and oxidized components (a conclusion from \citet{rubie2015accretion} supported here), Mars' mantle composition cannot be accurately reproduced if it is built from any significant fraction of material with a fully reduced iron component. 
Thus, in an Earth-producing protoplanetary disk, the vast majority of the mass of the Mars-like planet must originate in the outer more oxidized part of the disk, past $\delta$(1): exterior to  1.0 AU and (primarily) interior to 2.0 AU, as shown in Figures \ref{fig:idealtaylor} and \ref{fig:idealyoshi}. 
Some planetesimals that contribute a very small amount of mass to the Mars body may originate at even greater heliocentric distances, and in the case of Simulation 4:1-0.5-8, embryo 72 must originate at 2.68 AU in order to result in a Mars with \citet{taylor2013bulk} mantle composition.
When \citet{yoshizaki2020composition} is used as the composition of the Martian mantle, the region from which Mars can be sourced is even better constrained; nearly all the mass must originate between 1.0 and 1.3 AU, as shown in Figure \ref{fig:idealyoshi}. 

To first order, the different oxidation gradients necessary to reproduce the Martian mantle compositions according to \citet{taylor2013bulk} and \citet{yoshizaki2020composition} are due to their differences in estimated FeO content, as shown in Figures \ref{fig:idealtaylor} and \ref{fig:idealyoshi}.
Compared to the FeO content of Earth ($\sim$8.1 wt\% \citep{palme2014cosmochemical}), the Martian mantle is more oxidized regardless of which estimate is used. 
Accordingly, we find that the vast majority of Martian building blocks must come from the more oxidized exterior portion of the disk between 1.0 and 2.0 AU, regardless of which Martian mantle composition is being fit.
Between the two Martian mantle estimates, \citet{taylor2013bulk} has a more oxidized composition with higher FeO composition (18.14 wt\%)  so Mars must be sourced from more oxidized regions of the disk.
In contrast, \citet{yoshizaki2020composition} predicts a more reduced Martian composition with a slightly lower FeO composition (14.67 wt\%). 
This slightly lower FeO composition allows for more contribution from the more reduced part of the disk.
Across all 6 simulations, the building blocks come from a wider variety of semi-major axes when fitting the \citet{taylor2013bulk} composition than the \citet{yoshizaki2020composition}.
In either case, the material to build Mars cannot come from the innermost reduced portion of the disk. 

\section{Discussion}
While we were unable to reproduce Earth and Mars simultaneously, we instead have calculated that, based on its mantle composition, the building blocks of Mars must primarily come from an exterior region between 1.0 and 2.0 AU in the protoplanetary disk, contrasting with Earth where its building blocks have been found to likely originate from interior to 1.0 AU \citep[e.g.][]{rubie2015accretion,dauphas2017isotopic}. 
Encouragingly, this is consistent with isotopic evidence which suggests that Mars did not originate in the innermost portion of the Solar System \citep{brasser2017cool}.
This also confirms the results of \citet{rubie2015accretion} concluding again that the chemistry of the Mars is highly sensitive to its accretion history.

Our approach is based on assumptions that should be stated clearly.
First, we have assumed that Grand Tack simulations are broadly representative of terrestrial planet formation scenarios with strong radial mixing. 
This is reasonable because Grand Tack simulations are generally very successful in reproducing many of the features of the terrestrial planets and possess many of the same features of terrestrial planet formation as other scenarios with a significant giant impact phase.
In addition, the six N-body simulations used here are only meant as particular instances to show how the resulting Earth-like planets all have similar feeding zones but the Mars-like planets have widely varying feeding zones. 
With enough numerical realizations of the Grand Tack scenario, it may be possible to find many simulations that can simultaneously satisfy the constraints of both the terrestrial and Martian mantles.
Indeed, the Solar System could be a relatively rare outcome from a generic set of initial conditions, however we did not need to numerically reproduce this potentially rare outcome in order to assess the origin of the Martian building blocks in the inner disk.

Second, the assumed mantle composition of Mars is also a potential source of uncertainty. 
The small sampling size of Martian meteorites and the lack of ability to select samples more representative of mantle material increases the uncertainty of the composition of Mars' mantle. 

Third, in Tables 5-9, we show uncertainty estimates of mantle abundances calculated by standard error propagation from the formal 1-sigma partition coefficient errors associated with the fitting of the laboratory experiments \citep[similar to][]{rubie2015accretion}.
This procedure likely overestimates the uncertainties because standard error propagation assumes that the errors are independent, however the partition coefficients have highly correlated errors \citep[e.g.,][]{mann2009evidence}. 
Standard error propagation also assumes that the model behaves linearly, which it certainly does not, so this procedure provides only a first-order approximation.
Nevertheless, because our focus here is on a general trend resulting from the (well-documented) FeO-rich bulk composition of Mars, our fundamental thesis is unchanged even if we only consider mantle Fe abundance.

Lastly, in order to obtain our results, we manipulated the origin of the building blocks of Mars within an Earth-producing disk, guided by matching the FeO content of Mars-producing components to locations in the Earth-producing Fe-oxidation gradient as described in Section \ref{section:reconcile_model}.
A potential complication is that the initial Si oxidation state also varies as a function of heliocentric distance, as described in Section \ref{section:methods}, and so the FeO is not the sole determinant of the initial composition.
When fitting to the \citet{yoshizaki2020composition} Martian mantle, variation in Si is inconsequential because in the best-fit scenario for either the Mars-forming or Earth-forming disks, all the Martian building blocks originate from regions in the Mars-forming disk where Si is completely oxidized.
A small fraction ($\sim$ 10\%) of the mass that builds a Mars matching the mantle composition of \citet{taylor2013bulk} is reduced, however since the amount of reduced Si contributing to the Mars body is still small, the impact of this effect is limited and this does not affect the key point of our study, that Mars' mantle composition requires that its building blocks generally come from the more oxidized, exterior part of the inner disk. 

Our findings are intuitive, yet non-trivial; Mars must be built from material primarily sourced between 1.0 and 2.0 AU in order to achieve an accurate composition within an Earth-forming disk.
That Mars ought to come from a specific region in the disk matching its composition and moreover that this location would be near its present-day location in the disk seems intuitive.
However, the chaotic nature of planetary interactions during the period of accretion means that Mars' building blocks do not necessarily have to originate near Mars' present-day semi-major axis in the outer part of the inner Solar System.
Additionally, constraining Mars' dynamical formation history in this way would not be possible without examining its bulk chemical composition as we do here.

\section{Conclusions}

While some existing formation models can match the major astrophysical properties of our Solar System, thus far, none have succeeded in additionally simulating simultaneously the mantle chemistry of both Earth and Mars.  
We attempted to simultaneously model the formation of Earth and Mars within the Grand Tack scenario \citep{walsh2011low} using a joint astrophysical N-body and core-mantle differentiation simulation. 
Our model varies 5 free parameters characterizing the initial oxidation state in the protoplanetary disk and core formation pressures to find the starting conditions that best result in a simulated Earth or Mars with a mantle composition matching observed values. 
Our model was unable to simultaneously recreate the observed astrophysical properties and bulk chemistry of both Earth and Mars with satisfactory accuracy. 
When we instead used our model to reproduce Earth and Mars individually, we were successful.

We first used our model to reproduce Earth's composition and found the ideal Fe oxidation state of the initial protoplanetary disk for 6 different Grand Tack formation scenarios. 
Given that Earth has a large mass relative to the mass of the inner Solar System and it forms from components from the nearly the entirety of the inner Solar System, the best fit disk(s) we found to produce Earth is likely to be representative of the  oxidation state of the disk in most any traditional planetesimal accretion scenario. 
This finding matches the results in \citet{rubie2015accretion} successfully.
Next we used our model to reproduce Mars and found that, in the 6 simulations examined here, the initial oxidation state that produces a Mars-like planet is distinct from one that produces an Earth-like planet. 

This trend is robust; even if we found compatible initial oxidation gradients for Earth- and Mars-producing disks, this would be merely coincidental. 
Because Mars has a very small mass and its composition is more influenced by the composition of its individual components, the initial conditions to produce Mars are unlikely to ever be representative of true state of the disk as a whole. 
This general point will remain true, even with the use of higher resolution simulations. 
However, higher resolution simulations \citep[e.g.][]{walsh2019planetesimals, clement2020embryo, woo2021growing, woo2022terrestrial} still certainly have potential to reduce this stochasticity and improve constraints on the accretion history of Mars.

To reconcile the difference between a model protoplanetary disk that produces Earth with the disk that produces Mars, we determined where in the Earth-producing disk the Mars building blocks ought to originate.
We found that in order for Mars to form from the same protoplanetary disk that formed Earth, all of the mass that forms Mars must originate in the more oxidized region of the protoplanetary disk exterior to 1.0 AU. 
Indeed, \citet{rubie2015accretion} found that the final mantle FeO composition of simulated Mars bodies are highly influenced by the starting location of its initial embryo, a result that our study supports. 

Combining N-body and geochemical simulations as we do in this study can identify viable formation pathways that may account for the composition of our Solar System.
Here, we do not aim to pinpoint exactly where Mars formed, but instead present evidence that Mars' FeO-rich mantle chemistry could not be sourced from the innermost region of the protoplanetary disk: to produce the \citet{taylor2013bulk} estimate of Mars' mantle composition, nearly all Mars' building blocks must be sourced primarily between 1.0 and 2.0 AU to ensure that they are appropriately oxidized.
If the Martian mantle has the composition estimated by \citet{yoshizaki2020composition}, the region from which Mars may primarily originate is even more constrained: between 1.0 to 1.3 AU.
Our results are consistent with a rapid origin for Mars in the outer reaches of the inner Solar System and serve as a constraint for future attempts to determine potential Solar System formation scenarios.

\bibliographystyle{cas-model2-names}

\bibliography{manuscript_new}

\end{document}